# Fluctuating polytropic processes, turbulence, and heating

G. Livadiotis, D.J. McComas

Department of Astrophysical Sciences, Princeton University, Princeton, NJ 08544, USA; glivadiotis@princeton.edu

**Abstract**

This paper explores the thermodynamics of fluctuating polytropic processes and their connection to turbulence. It is shown that random fluctuations of polytropic processes produce a nonzero overall heating of a particle system, e.g., solar wind plasma flowing out through the heliosphere; while any nonturbulent heating can be thermodynamically described by typical nonfluctuating polytropic processes, turbulent heating can be thermodynamically described through fluctuating polytropic processes. First, we derive the expression of the overall process and find that polytropic fluctuations lead to heat entering the system even if the respective nonfluctuating process is adiabatic. The temperature of the solar wind plasma protons decreases with heliospheric distance less than the adiabatic cooling, again, similar to when heating enters the system; this subadiabatic cooling is proportional to the variance of the fluctuations. We derive the heliospheric radial profiles of the thermodynamic expressions of the polytropic index, temperature, and heating rates. Then, we show that the analytical profiles of heating of fluctuating polytropic processes and of turbulent heating are identical, suggesting that turbulence heats plasma particle populations by fluctuating their polytropic processes. We apply the thermodynamics of fluctuating polytropic processes to the energy transfer from pickup ions (PUIs) to solar wind plasma protons, and derive the analytical expressions of PUI turbulent and nonturbulent heating rates, which are well fitted to the respective observations. Finally, we apply the thermodynamic model to the radial profile of PUI energy transfer to the solar wind plasma protons, where we derive the portion of PUI turbulent vs. nonturbulent heating rates.

## 1. Introduction

Space plasma particle distributions are characterized by polytropic thermodynamic processes. Such an example is the solar wind plasma proton population that expands out through the heliosphere. The solar wind plasma protons expand in the inner heliosphere near 1 au under nearly adiabatic processes, while in general solar wind flows under nonadiabatic polytropic processes. This is true even though the most probable among all the processes is the adiabatic one. (For more details, see: e.g., G. Livadiotis & M.I. Desai 2016; G. Livadiotis 2018; 2021; G. Nicolaou & G. Livadiotis 2019; G. Nicolaou et al. 2020; M.A. Dayeh & G. Livadiotis 2022; C. Katsavrias et al. 2024; 2025).

Adiabaticity is a special case of polytropic processes where no heat enters or exits the particle system as it flows along. Heating is precisely determined by the relationship between the polytropic $\gamma$ and adiabatic $\gamma_a$ indices. When the polytropic index is smaller than the adiabatic index ($\gamma < \gamma_a$), the particle system is heated by an external source, while when the polytropic index is larger ($\gamma > \gamma_a$), the particle system becomes a source that heats some other external sink. In this paper, we derive the overall heat exchange of solar wind protons, which are subject to fluctuations of a polytropic process.

In the inner heliosphere near 1 au, where there is no significant source or sink of heating of solar wind plasma protons, the polytropic index is nearly adiabatic. Sources of solar wind heating include turbulence and pickup ions (PUIs). Alfvén-wave turbulence is a major mechanism for heating the solar wind in the innermost heliosphere at heliospheric distances much closer to the Sun, where the heating rate is ~4-5 orders of magnitude larger than near 1 au (e.g., S. Bourouaine et al. 2024). On the other hand, PUIs drive turbulence in the outer heliosphere, with radial distance, latitude, and solar cycle dependences (e.g., G. P. Zank et al. 2018; B. Wang et al. 2023). When interstellar neutrals drifting into the heliosphere are ionized, they become picked-up by the solar wind and immediately begin to spiral around the magnetic field lines carried by the solar wind, becoming part of the outflowing plasma (e.g., R. Schwenn & E. Marsch 1990; H. J. Fahr et al. 2003). But the PUI density and pressure are too low near 1au to drive substantial heating (e.g., G. P. Zank et al. 2018; G. Livadiotis & D.J. McComas 2025a).

Therefore, solar wind protons are nearly adiabatic around 1 au, consistent with the absence of a significant heating source. However, purely adiabatic processes don't actually exist since random fluctuations always perturb these processes around their average adiabatic process. We ask: *What is the effect of random fluctuations of polytropic processes on the overall heating?* Even when the fluctuations of polytropic indices are symmetric and normally distributed around the adiabatic index, it is not ensured that the overall heating will be zero. Namely, positive and negative fluctuations ($\delta\gamma = \gamma - \gamma_a$) may not have equal and opposite impacts on heating, and thus their overall effect may not cancel out. In order to examine the overall effect on heating, we calculate the heating rate and gradient of particle systems flowing under fluctuating polytropic processes, and then apply this theory to solar wind plasma protons.

The heating of a particle system depends on the difference between its polytropic index and the adiabatic index. However, the polytropic index does not have a constant value throughout the heliosphere, but instead has a slowly varying radial profile (e.g., G. Livadiotis 2021), which is related to the radial profile of heating (e.g., G. Livadiotis 2019a;b; G. Livadiotis & D.J. McComas 2023a). Indeed, a single polytropic index corresponds to an ideal polytropic process, called a "polytrope", but as we will see in this paper, in general, a variable polytropic index can be decomposed into a sum or integral of polytropes, that is, expressed as a superposition of multiple polytropes, or a "multi-polytrope."

The paper is organized as follows. In Section 2, we present the ideal polytropic processes (polytropes), while in Section 3, we introduce the concept of superposition of polytropic processes (multi-polytropes), presenting the concept of fractional effective degrees of freedom, their variability, and the superposition of those degrees of freedom once they are randomly distributed. In Section 4, we develop the superposition of polytropic processes with random fluctuations; we derive the generalized polytropic relationship $n(T)$, the polytropic index, $v(T)$ and $\gamma(T)$, the heating of polytropic processes. In Section 5, we apply the theory to the solar wind protons in the inner heliosphere, and derive the temperature, $T(r)$, the polytropic index, $v(r)$ and $\gamma(r)$, heating rate $dq(r)/dt$ and gradient $dq(r)/dr$. In Section 6, we show the connection of fluctuating polytropic processes with heliospheric turbulence. For this, we examine the radial profiles of ideal and fluctuating polytropic processes and compare with the radial profile of turbulent heating. As an application, we apply the thermodynamic model to the radial profile of PUI energy transfer to the solar wind plasma protons, where we derive the portion of PUI turbulent vs. nonturbulent heating rates. Finally, in Section 7, we summarize and discuss this work and provide conclusions, while the Appendices provide additional details on the analytical derivations in this paper.

**2. Ideal polytropic processes: Polytropes**

As solar wind plasma flows outward along a streamline, its thermodynamic properties change according to polytropic processes. In these processes, thermal pressure $p$ and density $n$ are related by a power-law equation, $p \propto n^\gamma$. Given the ideal gas equation of state that generally applies for space plasmas (e.g., F.J. Rogers 1994), $p = n k_B T$, a similar power-law relationship exists between density $n$ and temperature $T$, i.e., $n \propto T^v$. The primary $\gamma$ and secondary $v$ polytropic indices are related by

$$\gamma = 1 + 1/v \iff v = 1/(\gamma - 1). \qquad (1)$$

The polytropic index of solar wind plasma is, in general, not constant over long heliospheric distances. In the case of the solar wind protons, the further from the Sun, the more the polytropic index deviates from its adiabatic value. It decreases slowly from 1 to 5 au ($\Delta\gamma \sim -0.05$ per 1au; G. Livadiotis & D.J. McComas 2023a; G. Nicolaou et al. 2023), reaching an isothermal value ($\gamma = 1$) near ~ 30 au (H.A. Elliott et al. 2019), beyond which, it decreases further to sub-isothermal values ($\gamma < 1$) in the outer heliosphere (G. Livadiotis & D.J. McComas 2012; 2013; G. Livadiotis et al. 2022; G. Livadiotis et al. 2025).

In the case of variable polytropic processes, the polytropic index is defined by the proportional changes of pressure with respect to density, or, for the secondary polytropic index, by the proportional changes of density with respect to temperature,

$$\gamma = d \ln p / d \ln n, \quad v = d \ln n / d \ln T, \qquad (2)$$

which, again, is consistent to Eq.(1).

When the polytropic index is constant, the process is described by a single power-low exponent,

$$\Pi = n \cdot T^{-v}. \qquad (3)$$

Equation (3) describes ideal polytropic processes (polytropes), as opposed to the polytropic processes of variable polytropic index. While the value of $\Pi$ remains constant along a streamline, each streamline is characterized by a specific value of Π, which may be different amongst adjacent streamlines. Thus, the duplet {*T*,Π} constitutes a reparameterization of the thermodynamic phase space of {*T*,*n*}, a curvilinear grid of coordinates for describing the set of all streamlines with polytropic index *v* (demonstrated in Figure 1). Therefore, the polytropic relationship does not reduce the number of independent thermodynamic parameters (i.e., any two of the triplet *n*, *T*, *P*).

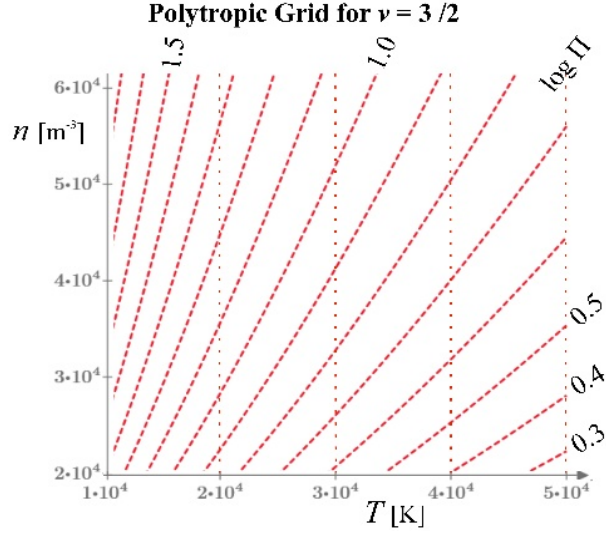

Figure 1. Thermodynamic grid: The two independent thermodynamic parameters of density and temperature form this grid in linear coordinates. The ensemble of streamlines of all the polytropic processes with some constant polytropic index represent a curvilinear system of coordinates, with the transform of {*T*,*n*}→{*T*,Π} given by the polytropic relationship in Eq.(3). The plot depicts the values of Π: log Π = 0.3, 0.4, …, 1.7, for the adiabatic polytropic index *v* = 3 / 2.

### 3. Superposition of ideal polytropic processes: Multi-polytropes

#### *3.1. Fractional effective dimensionality*

Here we discuss the effective degrees of freedom, which are defined through the polytropic processes. In particular, the secondary polytropic index equals half the degrees of freedom, $v = \tfrac{1}{2}D$; these polytropic degrees of freedom become the kinetic degrees of freedom in the case of an adiabatic index, $v_a = \tfrac{1}{2}D_K$. Using the polytropic indices or the degrees of freedom is equivalent; we may always switch from the nomenclature of polytropic and adiabatic indices, $\gamma = 1 + 2/D$ and $\gamma_a = 1 + 2/D_K$, to that of degrees of freedom, i.e., $D$ and $D_K$, respectively, where the degrees of freedom are more fundamental as they represent an actual, physically meaningful quantity. The dimensionality or degrees of freedom can be a fractional number as the velocity distribution often does not fully populate all the possible portions of the phase space represented by integer dimensionality. This is called effective dimensionality or effective degrees of freedom, as it characterizes thermodynamics in the same way as the traditionally assumed three kinetic degrees of freedom.

An important example of fractional effective dimensionality is the PUI population, whose distribution evolves from an initial 1D ring into a 3D filled sphere (e.g., S. V. Chalov & H. J. Fahr 1998; J. Y. Lu & G. P. Zank 2001; B. M. Randol et al. 2013; E. J. Zirnstein et al. 2022). In particular, the initial ring beam is unstable and drives kinetic instabilities (e.g., C. T. Russell et al. 1998; D. E. Huddleston et al. 2000; K. Min et al. 2017; S. Du et al. 2025). Immediately, the ring is transformed into a two-dimensional bispherical shell due to plasma waves that initiate pitch angle scattering (e.g., C. S. Wu & R. C. Davidson 1972; M. A. Lee & W. H. Ip 1987; H. J. Fahr 2007; E. J. Summerlin et al. 2014; V. Florinski et al. 2016; K. Min et al. 2018). With continuous scattering and advection in the expanding solar wind, the distribution eventually fills the entire velocity sphere, becoming fully isotropic and three-dimensional (e.g., S. V. Chalov & H. J. Fahr 1998; J. Y. Lu & G. P. Zank 2001; E. J. Zirnstein et al. 2022). Through the fractional dimensionality, instead of thinking of the evolution of PUI distribution as discrete jumps from 1D → 2D → 3D, we can better interpret the intermediate states physically by using a fractional dimension to represent the progress in populating an increasingly large fraction of the available phase space; or, according to G. Livadiotis & D. J. McComas (2023a;b), "Due to thermal fluctuations, this one-dimensional ring is actually a higher dimensional distribution with an effective PUI dimensionality dispersed between values of $D \sim 1$ and $D \sim 2$, which is lower dimensionality ($D < 3$) than the three-dimensional plasma ($D = 3$)." Furthermore, the fractional effective dimensionality of PUIs is a measure of the order of their distribution; the lower the $D_{PUI}$, the higher the order of the PUI distribution. When a solar wind proton undergoes a charge exchange with an interstellar neutral, the newly born PUI is picked-up by the frozen-in magnetic field that flows along with the solar wind, outward through the heliosphere. The newly born PUIs replace the missing charge and number of solar wind protons that were involved in the charge exchange. What is not matched, however, are the degrees of freedom because at each charge exchange a 3D solar wind proton is replaced by highly ordered PUI that has a lower number of degrees of freedom, $D_{PUI} < 3$. Consequently, the entropy of the solar wind protons decreases as they continue to charge-exchange with neutrals. This is an important process that decreases the entropy and alters the thermodynamics of the solar wind protons as they move out through the heliosphere.

PUIs are not the only population that is characterized by a fractional effective dimensionality. Any space plasma particles may have a fractional number of effective degrees of freedom. In the case of solar wind protons, a systematic variation of their effective dimensionality can be mostly caused by (i) the incorporation of newly born PUIs into solar wind (as described above), and (ii) the strong long-range interactions that correlates a large number of particles. Specifically, we have: (i) PUIs decrease the per particle degrees of freedom of the system of solar wind protons; the PUI dimensionality is $1 \lessapprox D_{PUI} \lessapprox 3$, with $D_{PUI} \sim 1$ corresponding to the ring distribution of new born PUIs, while mature PUIs correspond to higher dimensionality up to $D_{PUI} \sim 3$ corresponding to their thermalization; thus, the per particle solar wind effective dimensionality becomes $1 \lessapprox D_{\text{eff}} \lessapprox 3$ (G. Livadiotis & D.J. McComas 2023b). (ii) Long-range interactions induce strong correlations among particles binding together the degrees of freedom of each individual particle into a "cloud" of degrees of freedom; the degrees of freedom are indistinguishable and among $N$ particles correspond to an effective dimensionality higher than the traditional three degrees of freedom per particle, i.e., $3 \lessapprox D_{\text{eff}} \lessapprox 3N$ (G. Livadiotis et al. 2024pui). Other factors may also vary the effective dimensionality of protons, such as the interplanetary magnetic field (Cuesta et al. 2025mag), but in addition to systematic variations that may exist or not, random fluctuations always characterize the variability of degrees of freedom.

Random fluctuations are intrinsic to systems composed of many particles. Even in thermodynamic equilibrium, physical quantities of particles fluctuate around mean values. Random fluctuations are often normally distributed. According to the Central Limit Theorem, the normal distribution is an attractor of statistical distributions. Namely, when many small and independent random effects contribute to a measurement, the resulting distribution converges toward a normal distribution (e.g., A.M. Selvam 2017).

Systematic or random variations of the degrees of freedom lead to more complex forms of the polytropic relationship between thermodynamic parameters than the typical power-law characterizing single polytropes. Any complex form of polytropic process can be equivalently expressed by a power law such as $p \propto n^\gamma$ or $n \propto T^\nu$, where the involved polytropic index is not a constant but it varies (e.g., with temperature, $\gamma(T)$ or $\nu(T)$). Therefore, a nonconstant polytropic index is consistent with variations of the effective degrees of freedom.

Next, we develop the method of decompose any functional polytropic relationship into the summation or integration of multiple polytropic processes, each characterized by a constant polytropic index or degrees of freedom. Then, we will apply this method to the case of normally distributed polytropic processes.

### 3.2. Superposition of degrees of freedom

A polytrope is described by the polytropic relationship of a single polytropic index, and it can be expressed in terms of a single value of the degrees of freedom as

$$n / n_* = \Pi \cdot (T / T_*)^{\frac{1}{2}D}, \tag{4}$$

where $\{T_*, n_*\}$ are characteristic scales of the system, universal for all the streamlines of the system; (as shown from Eq.(4), $\{T_*, n_*\}$ represents a point only on the certain streamline with $\Pi=1$).

The variation on the effective degrees of freedom $D$ has an overall effect on the thermodynamics of space plasmas, and specifically on their polytropic processes. All the different values of the degrees of freedom contribute to the overall polytropic process, described by summing over all the variable $D$. At any snapshot of the system in time, the degrees of freedom $D$ vary among the particles, described by a stationary statistical distribution $P_D$; this is similar to the variability of the energy or other physical quantities among the particles, described by stationary statistics and characterize the thermodynamics of the particle system.

As an example, let a particle population described by the discrete set of degrees of freedom {$D_1$, $D_2$, …, $D_w$ }, where the portion of particles in the system with degrees of freedom equal to $D_i$ is $P_{Di}$. Particles are undistinguishable, thus at any snapshot of particles, all the variable degrees of freedom $D_1$, $D_2$, …, $D_w$ coexist, having probabilities {$P_{D1}$, $P_{D2}$, …, $P_{Dw}$}, respectively. Each portion has a certain density $n_D$ and a polytropic process. For a given temperature $T$, the density of particles following the polytropic process with $D$ degrees of freedom is $n_D / n_* = \Pi_D \cdot (T/T_*)^{\frac{1}{2}D}$. Then, using the probability distribution of the degrees of freedom, we can derive the expectation value of any physical expression that involves the degrees of freedom, such as, the expectation value of the polytropic processes,

$$n / n_* = \sum_D P_D \cdot n_D / n_* = \Pi \cdot \sum_D P_D \cdot (T/T_*)^{\frac{1}{2}D}. \tag{5a}$$

Similarly, for the continuous description of variability of $D$, we have

$$n / n_* = \int P(D) \cdot n(D) / n_* \, dD = \Pi \cdot \int P(D) \cdot (T/T_*)^{\frac{1}{2}D} \, dD. \tag{5b}$$

Therefore, the overall density is given as a superposition performed over the whole set of the effective degrees of freedom. In general, we express this polytropic process using the probability distribution of the degrees of freedom (statistical weights of each $D$), either in a discrete $P_D$ or continuous $P(D)$ description, respectively:

$$\sum_D P_D = 1 \, , \, \int P(D) dD = 1. \tag{6}$$

When there is a single degree of freedom, then, the distribution becomes a delta of Kronecker's or a delta function (for the discrete/continuous descriptions, respectively), thus Eqs.(5a,b) recover the single polytrope, $\Pi = (n/n_*) \cdot (T/T_*)^{-\frac{1}{2}D}$.

Any (analytical) function can be expressed as a summation of powers forming the Maclaurin series, or as an integration on powers forming the Laplace transformation. Consequently, any functional polytropic relationship can be analyzed into a summation or integration of polytropes, according to Eq.(5a) and Eq.(5b), respectively. The addition or integration of all the polytropes form a multi-polytrope. Therefore, any multi-polytropic functional relationship can be analyzed into a superposition of simple polytropes.

## 4. Superposition of polytropic fluctuations
### *4.1. Generalized polytropic relationship n(T)*
4.1.1. Fluctuations of polytropic processes

Let a particle system flowing along a streamline from $(T_0, n_0)$ to $[T(\vec{r}), n(\vec{r})]$ under a multi-polytropic process described in Eq.(5b), i.e.,

$$n = n_0 \cdot \left[ \int P(D) \cdot (T_0 / T_*)^{\frac{1}{2}D} dD \right]^{-1} \cdot \int P(D) \cdot (T / T_*)^{\frac{1}{2}D} dD . \tag{7}$$

where the polytropic pressure $\Pi = (n_0 / n_*) \cdot \left[ \int P(D) \cdot (T_0 / T_*)^{\frac{1}{2}D} dD \right]^{-1}$ remains constant along a streamline.

The polytropic processes are characterized by fluctuations of the effective degrees of freedom having distribution $P(D)$, ensemble average $\bar{D} \equiv \langle D \rangle = \int D \cdot P(D) dD$ and standard deviation $\sigma^2 = \int (D - \bar{D})^2 \cdot P(D) dD$.

Next, we derive the generalized polytropic relationship that characterizes the superposition of the fluctuating polytropes. First, we work with any distribution $P(D)$ but in the approximation of small σ. Thereafter, we specifically use the formalism of normally distributed fluctuations, where we end up with the same generalized polytropic relationship, but this time the expression is exact.

4.1.2. Generalized polytropic process for any $P(D)$.

We observe that the superposition of $(T / T_*)^{\frac{1}{2}D}$ on $D$ leads to its ensemble average,

$$n(T) = n_0 \cdot \langle (T / T_*)^{\frac{1}{2}D} \rangle / \langle (T_0 / T_*)^{\frac{1}{2}D} \rangle . \tag{8}$$

The polytropic index is similarly formulated, i.e.,

$$\nu = \frac{d \ln n(T)}{d \ln T} = \frac{\langle \frac{1}{2} D \cdot (T / T_*)^{\frac{1}{2}D} \rangle}{\langle (T / T_*)^{\frac{1}{2}D} \rangle} . \tag{9}$$

In the approximation of small σ, we find $\langle e^{\frac{1}{2}\delta D \ln(T / T_*)} \rangle \cong 1 + \frac{1}{8} \ln^2 (T / T_*) \cdot \sigma^2$ and $\langle \frac{1}{2} \delta D \cdot e^{\frac{1}{2}\delta D \ln(T / T_*)} \rangle \cong \frac{1}{4} \ln^2 (T / T_*) \cdot \sigma^2$, where we set $\delta D \equiv D - \bar{D}$, thus we have $\langle \delta D \rangle = 0$ and $\sigma^2 \equiv \langle \delta D^2 \rangle$.

Hence, we find that the generalized polytropic relationship is given by the convex parabolic form

$$\ln \left( \frac{n}{n_0} \right) \cong \left[ \frac{1}{2} \bar{D} + \frac{1}{4} \ln \left( \frac{T_0}{T_*} \right) \sigma^2 \right] \cdot \ln \left( \frac{T}{T_0} \right) + \frac{1}{8} \sigma^2 \cdot \ln^2 \left( \frac{T}{T_0} \right), \tag{10}$$

while the polytropic index is

$$\nu(T) \cong \frac{1}{2} \bar{D} + \frac{1}{4} \ln^2 (T / T_*) \cdot \sigma^2 . \tag{11}$$

4.1.3. Generalized polytropic process for normally distributed fluctuations.

The effective degrees of freedom are characterized by random fluctuations, normally distributed around their mean $\bar{D}$ with a standard deviation σ, i.e.,

$$P(D) = \frac{1}{\sqrt{2\pi}\sigma} \cdot e^{-\frac{(D - \bar{D})^2}{2\sigma^2}} . \tag{12}$$

The superposition is formulated according to Eq.(5b). Then, the density is given by

$$n(T) = n_0 \cdot \tilde{F}(T) / \tilde{F}(T_0) \text{ with } \tilde{F}(T) \equiv \frac{1}{\sqrt{2\pi}\sigma} \cdot \int_{-\infty}^{+\infty} e^{\frac{1}{2}D \ln(T / T_*) - \frac{(D - \bar{D})^2}{2\sigma^2}} dD . \tag{13}$$

The exponent is rewritten after some calculus as

$$-\frac{\left[D-\bar{D}-\frac{1}{2}\sigma^2\ln(T/T_*)\right]^2}{2\sigma^2}-\frac{1}{2}\bar{D}\ln(T/T_*)-\frac{1}{8}\sigma^2\ln^2(T/T_*). \tag{14a}$$

The first component is equivalent to a normal distribution over $\bar{D}-\frac{1}{2}\sigma^2\ln(T/T_0)$, and is integrated as

$$\frac{1}{\sqrt{2\pi}\sigma}\cdot\int_{-\infty}^{+\infty}e^{-\frac{\left[D-\bar{D}-\frac{1}{2}\sigma^2\ln(T/T_*)\right]^2}{2\sigma^2}}dD=1, \tag{14b}$$

while the other two components are constants remaining out of the integration and leading to

$$\frac{n(\vec{r})}{n_0}=\left[\frac{T(\vec{r})}{T_*}\right]^{\frac{1}{2}\bar{D}+\frac{1}{8}\sigma^2\ln[T(\vec{r})/T_*]}\cdot\left(\frac{T_0}{T_*}\right)^{-\frac{1}{2}\bar{D}-\frac{1}{8}\sigma^2\ln(T_0/T_*)}, \tag{15a}$$

where we include the positional dependence along streamlines. On logarithmic scales (log $T$ and log $n$) the expression again becomes the convex parabola of Eq.(10), but in an exact formulation rather than an approximation,

$$\ln\left[\frac{n(\vec{r})}{n_0}\right]=\left[\frac{1}{2}\bar{D}+\ln\left(\frac{T_0}{T_*}\right)\frac{1}{4}\sigma^2\right]\cdot\ln\left[\frac{T(\vec{r})}{T_0}\right]+\frac{1}{8}\sigma^2\cdot\ln^2\left[\frac{T(\vec{r})}{T_0}\right]. \tag{15b}$$

4.1.4. Final expression

The convex parabola of Eqs.(15a,b) provides the generalized polytropic relationship $n(T)$ that characterizes fluctuating polytropic processes. Expressing this relationship in (log $T$, log $n$), we obtain

$$\ln[n(\vec{r})]=A_0+A_1\ln T(\vec{r})+A_2\ln^2 T(\vec{r}), \tag{16a}$$

where

$$A_0=\ln n_0-\frac{1}{2}\bar{D}\ln T_0-\sigma^2\frac{1}{4}\ln T_0(\ln T_*+2\ln T_0),\ A_1=\frac{1}{2}\bar{D}-\sigma^2\frac{1}{4}\ln T_*,\ A_2=\frac{1}{8}\sigma^2. \tag{16b}$$

The corresponding relationship in terms of density is approximated in terms of small $\sigma$ as,

$$\frac{n(\vec{r})}{n_0}=\left[\frac{T(\vec{r})}{T_0}\right]^{\frac{1}{2}\bar{D}}+\sigma^2 g(\vec{r}),\ g(\vec{r})=\frac{1}{4}\left[\frac{T(\vec{r})}{T_0}\right]^{\frac{1}{2}\bar{D}}\ln\left[\frac{T(\vec{r})}{T_0}\right]\left\{\frac{1}{2}\ln\left[\frac{T(\vec{r})}{T_0}\right]+\ln\left(\frac{T_0}{T_*}\right)\right\}. \tag{17}$$

The degrees of freedom of each particle is a random variable $D_i$ (Figure 2a), but over a larger number of particles or many samplings of a single particle, the observed values $D_i$ converge to the ensemble distribution $P(D)$ (Figure 2b). In Figure 2(c) the timeseries of the normally distributed random values of degrees of freedom, $D_i$, are used to derive the respective timeseries of densities through $(n_i/n_*)=\Pi\cdot(T/T_*)^{\frac{1}{2}D_i}$ (with $\Pi=10$ and $T_*=1$); then, the derived densities are binned over $\Delta N=150$ values to mimic the superposition of all the variable degrees of freedom. The resulting polytropic relationship is given by the concave parabola of Eq.(15b).

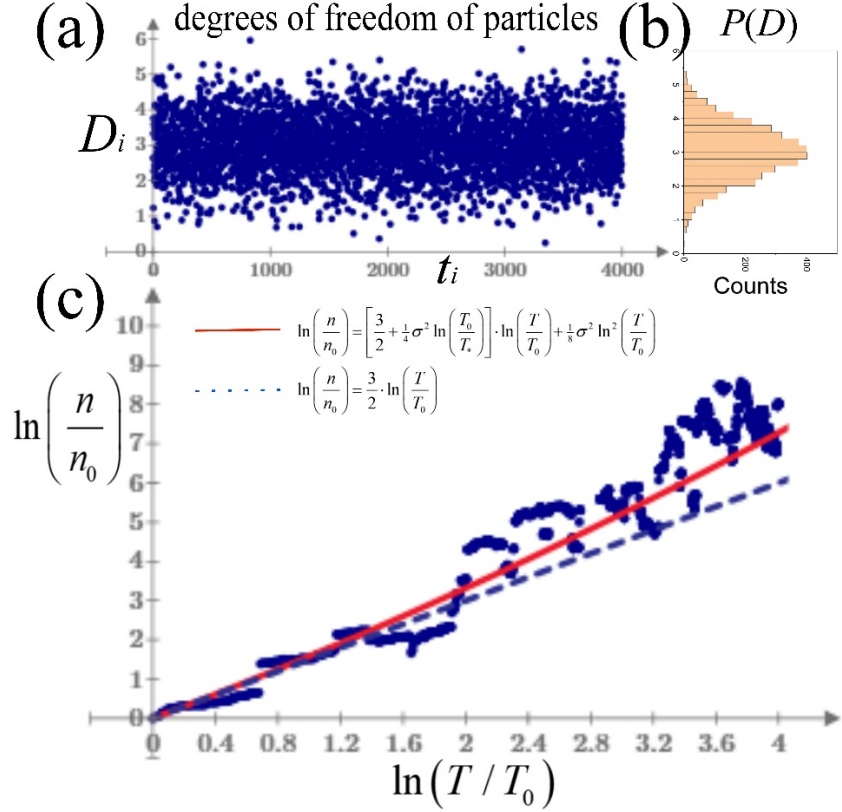

**Figure 2.** Generalized polytropic relationship for a simulation of randomly distributed degrees of freedom.

### 4.2. Polytropic index, ν(T) and γ(T)

The secondary polytropic index is given by $\nu(T) \equiv d \ln n / d \ln T$, which is the slope of the polytropic relationship, Eq.(15b), i.e.,

$$\nu(T) = \tfrac{1}{2}\bar{D} + \sigma^2 \cdot \tfrac{1}{4}\ln(T/T_*) \;,\; \gamma(T) = 1 + [\tfrac{1}{2}\bar{D} + \sigma^2 \cdot \tfrac{1}{4}\ln(T/T_*)]^{-1}. \tag{18}$$

The (half) mean effective degrees of freedom define the secondary index of a nonfluctuating polytropic process, $\nu_{\sigma=0} = \tfrac{1}{2}\bar{D}$, thus, the corresponding primary polytropic index is $\gamma_{\sigma=0} = 1 + 2/\bar{D}$. The secondary index of a nonfluctuating adiabatic process is $\nu_a = \tfrac{1}{2}D_a$, thus the corresponding primary polytropic index is $\gamma_a = 1 + 2/D_a$. Hence, the deviations of the polytropic indices from their adiabatic value are given by

$$\nu(T) - \nu_a = \nu_{\sigma=0} - \nu_a + \sigma^2 \cdot \tfrac{1}{4}\ln(T/T_*), \tag{19a}$$

$$\gamma_a - \gamma(T) = \gamma_a - \gamma_{\sigma=0} + \frac{\sigma^2 \cdot \tfrac{1}{4}(\gamma_{\sigma=0}-1)^2 \ln(T/T_*)}{1+\sigma^2 \cdot \tfrac{1}{4}(\gamma_{\sigma=0}-1)^2 \ln(T/T_*)}, \tag{19b}$$

with the ratio

$$\frac{\gamma_a - \gamma(T)}{\gamma_a - 1} = 1 - \frac{D_a}{\bar{D}} \cdot \left[1 + \tfrac{1}{4}\sigma^2 \cdot \ln(T/T_*)/\tfrac{1}{2}\bar{D}\right]^{-1}. \tag{19c}$$

The polytropic deviation $\gamma_a - \gamma(T)$ can be written for small values of sigma as $\gamma_a - \gamma(T) \cong \gamma_a - \gamma_{\sigma=0} + \sigma^2 \cdot \tfrac{1}{4}(\gamma_{\sigma=0}-1)^2 \ln(T/T_*)$. It is clear now that the deviation of a fluctuating polytropic process from the ideal adiabatic process, $\gamma_a - \gamma(T)$, is caused by two factors, (i) the deviation of the nonfluctuating (ideal) polytropic process from the ideal adiabatic one, expressed by the difference in their indices, $\gamma_a - \gamma_{\sigma=0}$, and (ii) the fluctuations of the polytropic process characterized by the variance $\sigma^2$. In the case of a fluctuating adiabatic process, the deviation from the ideal adiabatic process is simply caused by the fluctuations. The absolute difference between the adiabatic and fluctuating adiabatic indices increase with the variance $\sigma^2$. Indeed, we have $\nu_{\sigma=0} = \nu_a = \frac{1}{2}\overline{D} = \frac{1}{2}D_a$, thus

$$\nu(T) - \nu_a = \sigma^2 \cdot \tfrac{1}{4} \ln(T/T_*), \qquad (20a)$$

and

$$\gamma_a - \gamma(T) = \frac{\sigma^2 \cdot \tfrac{1}{4}(\gamma_a - 1)^2 \ln(T/T_*)}{1 + \sigma^2 \cdot \tfrac{1}{4}(\gamma_a - 1)\ln(T/T_*)} \cong \sigma^2 \cdot \tfrac{1}{4}(\gamma_a - 1)^2 \ln(T/T_*) + O(\sigma^4) \; . \qquad (20b)$$

When the nonfluctuating processes are adiabatic, $\gamma_{\sigma=0} = \gamma_a$, they become subadiabatic in the presence of fluctuations. In particular, the primary polytropic index becomes smaller with fluctuations, $d\gamma/d\sigma < 0$. When the nonfluctuating processes are superadiabatic, $\gamma_{\sigma=0} > \gamma_a$, the resulted polytropic index is closer to the adiabatic value, and even more, for sufficiently large $\sigma$, it can be turned into subadiabatic. In particular, the standard deviation must be $\sigma > \sigma_c$, where $\sigma_c \equiv 2\sqrt{(\nu_a - \nu_{\sigma=0})/\ln(T/T_*)}$, for the resulted polytropic index to be subadiabatic, $\gamma(T) < \gamma_a$.

The value of the polytropic index is related to the heat entering or exiting the system. When the primary polytropic index is smaller than the adiabatic one, $\gamma < \gamma_a$, the particle system is heated, while when is larger, $\gamma > \gamma_a$, heat is exiting the system. The heating rate is proportional to the difference of adiabatic and polytropic indices (e.g., G. Livadiotis & D.J. McComas 2023a). Therefore, the stronger the fluctuations, that is, the larger the variance of the normally distributed fluctuations, the larger the heating of the system. In the next section, we derive the exact formulation of the heating and its dependence on the variance of fluctuations, $\sigma^2$.

### *4.3. Heating of polytropic processes*

For expressing heating in terms of the polytropic index, we revisit the analytical procedure in the work of G. Livadiotis & D.J. McComas (2023a; 2025a) (see also: Livadiotis 2019a; G. Livadiotis & D.J. McComas 2023c). According to this, the first law of thermodynamics states that heating $dQ$ is shared between changes in the internal energy $dU$ and the work produced $dW$. For a system of $N$ particles and $D_K$ kinetic degrees of freedom per particle the internal energy varies by $dU = \tfrac{1}{2}D_K N k_B dT$, while from expansion or contraction of the system's volume by $dV$, the work varies by $dW = pdV$, that is,

$$dQ = dU + dW = \tfrac{1}{2}D_K(pdV + Vdp) + pdV = (\tfrac{1}{2}D_K + 1)pdV + \tfrac{1}{2}D_K Vdp \; . \qquad (21)$$

Using $pV = Nk_B T$ and dividing by $\tfrac{1}{2}D_K pdV$, we obtain:

$$(1+\tfrac{2}{D_K}) + \frac{d\ln p}{d\ln V} = \frac{dq}{\tfrac{1}{2}D_K k_B T d\ln V}, \qquad (22)$$

where we used $q \equiv Q/N$ as the per-particle heating of the system. Equation (9) is written in terms of the polytropic and adiabatic indices, $\gamma_a \equiv 1 + 2/D_K$ and $\gamma \equiv d\ln p/d\ln n = -d\ln p/d\ln V$ or $\Leftrightarrow$ $d\ln n/d\ln T = \gamma - 1$, leading to

$$\frac{\gamma_a - \gamma}{\gamma_a - 1} = \frac{dq}{k_B T \cdot d\ln V} . \qquad (23)$$

We observe that heating is proportional to the difference between the adiabatic and polytropic indices. Solving in terms of the heating rate, we obtain the radial gradient and rate of the heating to be respectively given by:

$$\frac{dq}{dr} = \frac{\gamma_a - \gamma}{\gamma_a - 1} \cdot k_B T \cdot \frac{d\ln V}{dr} \text{ and } \frac{dq}{dt} = u_b \cdot \frac{dq}{dr}, \qquad (24)$$

introducing the bulk speed of the flow, $u_b$. (Another way of characterizing heating is through the heat capacity; see the respective formulation in Appendix A). In order to express the radial profile of density, we use the conservation of flux, $n(r)u_b(r)r^2 = n(r_0)u_b(r_0)r_0^2 : const.$, setting $r_0 = 1$ au, and the radial exponents of density $n(r) \sim r^{-a_n}$ and of speed $u_b(r) \sim r^{-a_u}$, or, $-a_n - a_u + 2 = 0$, with $a_n \sim 1.8 - 1.9$ and $a_u \sim 0.1$ (representing the loss of solar wind protons through charge exchange with interstellar neutrals, G. Livadiotis & D.J. McComas 2025b). Therefore, we have $d\ln V = -d\ln n = a_n d\ln r = (a_n/r)dr = (a_n u_b/r)dt$, and thus, the heating gradient and rate in Eq.(24) become

$$\frac{dq}{dr} = \frac{\gamma_a - \gamma}{\gamma_a - 1} \cdot a_n k_B T / r \text{ and } \frac{dq}{dt} = u_b \cdot \frac{dq}{dr} . \qquad (25)$$

We have already found the polytropic relationship for the fluctuating polytropic processes, e.g., see Eq.(19c). This is used in Eq.(25) to find the (logarithmic) gradient of heating,

$$\begin{aligned}\frac{d}{d\ln r}\left[\frac{q(x)}{k_B T_*}\right] &= a_n \cdot \left[1 - \frac{D_a}{\bar{D}} \cdot (1 + \sigma^2 \cdot \tfrac{1}{4}\ln x / \tfrac{1}{2}\bar{D})^{-1}\right] \cdot x \\ &\cong a_n \cdot \left(1 - \frac{D_a}{\bar{D}}\right) \cdot x + \sigma^2 \cdot \tfrac{1}{2} a_n \frac{D_a}{\bar{D}^2} \cdot x\ln x \\ &= a_n \frac{\gamma_a - \gamma_{\sigma=0}}{\gamma_a - 1} \cdot x + \sigma^2 \cdot \frac{a_n(\gamma_{\sigma=0} - 1)^2}{4(\gamma_a - 1)} \cdot x\ln x,\end{aligned} \qquad (26)$$

where we used the identity $(\gamma_a - \gamma_{\sigma=0})/(\gamma_a - 1) = 1 - D_a/\bar{D}$ and the definition $x = T/T_*$. The approximation for small $\sigma$ shows clearly that the heating is the result of two factors, (a) the nonfluctuating term, which is caused by the difference between adiabatic index and the polytropic index of the nonfluctuating process, and (b) the fluctuations of the polytropic process, proportional to $\sigma^2$. Observations that present a deviation of heating from linearity with respect to temperature, especially if this deviation is proportional to the logarithm of temperature, would be a strong indication of the effect of fluctuations:

$$\frac{1}{k_{\mathrm{B}} T} \cdot \frac{dq}{d \ln r} \cong a_n \cdot \frac{\gamma_a - \gamma_{\sigma=0}}{\gamma_a - 1} + \sigma^2 \cdot \frac{a_n (\gamma_{\sigma=0} - 1)^2}{4(\gamma_a - 1)} \cdot \ln(T / T_*) , \quad (27a)$$

for instance, setting $D_a = 3$ and $a_n = 2$, we have

$$\frac{1}{k_{\mathrm{B}} T} \cdot \frac{dq}{d \ln r} \cong 3 \cdot (5/3 - \gamma_{\sigma=0}) + \sigma^2 \cdot \tfrac{3}{4} (\gamma_{\sigma=0} - 1)^2 \cdot \ln(T / T_*) . \quad (27b)$$

Again, we clearly observer that the heating is caused by two factors, (i) the difference between the nonfluctuating (ideal) polytropic and the ideal adiabatic indices or degrees of freedom, $\bar{D} - D_a \neq 0$, and (ii) the variance $\sigma^2$ that measures the fluctuations of the polytropic process.

## 5. Heliospheric radial profiles of thermodynamic parameters

In this section we derive the radial profiles of thermodynamic quantities of the solar wind protons in the inner heliosphere, considering that the plasma protons flow radially outward under a fluctuating polytropic process. We find the analytical expressions of temperature, polytropic index, and heating.

### 5.1. Temperature, T(r)

We first invert the polytropic relationship in Eq.(15b) and express the temperature in terms of the density. For this, we solve the binomial

$$\tfrac{1}{8} \sigma^2 \ln^2 \left[ \frac{T(\vec{r})}{T_0} \right] + \left[ \tfrac{1}{2} \bar{D} + \tfrac{1}{4} \sigma^2 \ln \left( \frac{T_0}{T_*} \right) \right] \cdot \ln \left[ \frac{T(\vec{r})}{T_0} \right] - \ln \left[ \frac{n(\vec{r})}{n_0} \right] = 0 , \quad (28a)$$

leading to

$$\left( \frac{\sigma}{\bar{D}} \right)^2 \tfrac{1}{2} \bar{D} \ln \left[ \frac{T(\vec{r})}{T_*} \right] = -1 + \sqrt{\left[ 1 + \left( \frac{\sigma}{\bar{D}} \right)^2 \tfrac{1}{2} \bar{D} \ln \left( \frac{T_0}{T_*} \right) \right]^2 + 2 \left( \frac{\sigma}{\bar{D}} \right)^2 \ln \left[ \frac{n(\vec{r})}{n_0} \right]} . \quad (28b)$$

(Note that only positive root gives finite solution for $\sigma$ tending to zero.) The expression is expanded in orders of $\sigma^2$,

$$\tfrac{1}{2} \bar{D} \ln \left[ \frac{T(\vec{r})}{T_0} \right] \cong \ln \left[ \frac{n(\vec{r})}{n_0} \right] - \tfrac{1}{2} (\sigma / \bar{D})^2 \cdot I(\vec{r}) , \quad (29a)$$

where

$$I(\vec{r}) = \ln \left[ \frac{n(\vec{r})}{n_0} \right] \cdot \left\{ \ln \left[ \frac{n(\vec{r})}{n_0} \right] + \bar{D} \ln \left( \frac{T_0}{T_*} \right) \right\} . \quad (29b)$$

Moreover, we approach the density expression with $n(r) = n_0 \cdot (r / r_0)^{-a_n}$, as in Eqs.(24,25). Then, we derive the radial profile of temperature

$$\left( \frac{\sigma}{\bar{D}} \right)^2 \tfrac{1}{2} \bar{D} \ln \left[ \frac{T(r)}{T_*} \right] = -1 + \sqrt{\left[ 1 + \left( \frac{\sigma}{\bar{D}} \right)^2 \tfrac{1}{2} \bar{D} \ln \left( \frac{T_0}{T_*} \right) \right]^2 - \left( \frac{\sigma}{\bar{D}} \right)^2 2 a_n \ln \left( \frac{r}{r_0} \right)} , \quad (30a)$$

approximated for small $\sigma$ to the parabolic logarithmic expression

$$\ln \left[ \frac{T(r)}{T_0} \right] \cong -\frac{2 a_n}{\bar{D}} \ln \left( \frac{r}{r_0} \right) - \sigma^2 \cdot \frac{a_n}{\bar{D}^2} \ln \left( \frac{r}{r_0} \right) \cdot \left[ \frac{a_n}{\bar{D}} \ln \left( \frac{r}{r_0} \right) - \ln \left( \frac{T_0}{T_*} \right) \right] , \quad (30b)$$

which reduces the power-law-like expression,

$$T(r) \cong T_0 \cdot \left(\frac{r}{r_0}\right)^{-a_T}, \tag{31a}$$

with a radially dependent exponent

$$a_T(r) = a_{T,0} + \sigma^2 \cdot a_{T,1}(r) \text{ with } a_{T,0} = \frac{2a_n}{\bar{D}} \text{ and } a_{T,1} = \frac{a_n}{\bar{D}^2} \cdot \left[\frac{a_n}{\bar{D}} \ln\left(\frac{r}{r_0}\right) - \ln\left(\frac{T_0}{T_*}\right)\right]. \tag{31b}$$

We observe that the radial exponent $a_T$ has two terms, the constant term that stands for the nonfluctuating polytropic process, $T \propto r^{-2a_n/\bar{D}}$, and the $\sigma^2$ term that has a logarithmic radial dependence. For the typical cases of $a_n \cong 2$ and the adiabatic process $\bar{D} \cong 3$, we obtain the known expression of adiabatic cooling for the first term, $T(r) \cong T_0 \cdot r^{-a_{T,0}}$, $a_{T,0} = 4/3$ while the second term, $r^{-s\sigma^2 \cdot a_{T,1}}$, $a_{T,1} = (4/27)\left[\ln(r/r_0) - (3/2)\ln(T_0/T_*)\right]$, expresses the larger contribution to the nonadiabatic cooling that can be either increasing the cooling, if $\ln(r/r_0) - (3/2)\ln(T_0/T_*) > 0$, or reducing it, if $\ln(r/r_0) - (3/2)\ln(T_0/T_*) < 0$. In order to find the exact expression of the heating, we need first to derive the polytropic index radial profile.

### 5.2. Polytropic index, ν(r) and γ(r)

Having found the profile of temperature in Eq.(30a), we use it in the expression of polytropic index as a function of temperature, Eq.(20a,b); the substitution leads to the radial profile of polytropic index. This derivation becomes straightforward when we observe that there is an argument $G_\sigma$, defined by

$$G_\sigma(r) \equiv 1 + \left(\frac{\sigma}{\bar{D}}\right)^2 \tfrac{1}{2}\bar{D}\ln\left[\frac{T(r)}{T_*}\right], \quad G_{\sigma=0} = 1, \tag{32a}$$

which is involved in all the thermodynamic variables we are interested in finding. Namely, for the polytropic index expressions,

$$\nu(r) = \tfrac{1}{2}\bar{D} \cdot G_\sigma(r), \quad \gamma(r) = 1 + (\tfrac{1}{2}\bar{D})^{-1} \cdot G_\sigma(r)^{-1}, \quad \frac{\gamma_a - \gamma(r)}{\gamma_a - 1} = 1 - \frac{D_a}{\bar{D}} \cdot G_\sigma(r)^{-1}, \tag{32b}$$

and the expression of temperature's radial profile,

$$1 + \left(\frac{\sigma}{\bar{D}}\right)^2 \tfrac{1}{2}\bar{D}\ln\left[\frac{T(r)}{T_*}\right] = G_\sigma(r) = \sqrt{\left[1 + \left(\frac{\sigma}{\bar{D}}\right)^2 \tfrac{1}{2}\bar{D}\ln\left(\frac{T_0}{T_*}\right)\right]^2 - \left(\frac{\sigma}{\bar{D}}\right)^2 2a_n \ln\left(\frac{r}{r_0}\right)}. \tag{32c}$$

Therefore, we find the radial profiles of polytropic index by substituting Eq.(32a) into Eq.(32b), i.e.,

$$\nu(r) = \tfrac{1}{2}\bar{D} \cdot \sqrt{\left[1 + \left(\frac{\sigma}{\bar{D}}\right)^2 \tfrac{1}{2}\bar{D}\ln\left(\frac{T_0}{T_*}\right)\right]^2 - \left(\frac{\sigma}{\bar{D}}\right)^2 2a_n \ln\left(\frac{r}{r_0}\right)}, \tag{33a}$$

$$\gamma(r) = 1 + (\tfrac{1}{2}\bar{D})^{-1} \cdot \left\{\left[1 + \left(\frac{\sigma}{\bar{D}}\right)^2 \tfrac{1}{2}\bar{D}\ln\left(\frac{T_0}{T_*}\right)\right]^2 - \left(\frac{\sigma}{\bar{D}}\right)^2 2a_n \ln\left(\frac{r}{r_0}\right)\right\}^{-\tfrac{1}{2}}. \tag{33b}$$

For small sigma, the argument $G_\sigma(r)$ is approximated by expanding it up to the 4th term,

$$G_\sigma(r) \cong 1 + g_2 \left(\frac{\sigma}{\bar{D}}\right)^2 + g_4 \left(\frac{\sigma}{\bar{D}}\right)^4 + O\left(\frac{\sigma}{\bar{D}}\right)^6, \text{ with} \tag{34a}$$

$$g_2 = \tfrac{1}{2}\bar{D}\left[\ln\left(\frac{T_0}{T_*}\right) - \frac{2a_n}{\bar{D}}\ln\left(\frac{r}{r_0}\right)\right], \quad g_4 = \tfrac{1}{2}\bar{D}a_n\ln\left(\frac{r}{r_0}\right)\left[\ln\left(\frac{T_0}{T_*}\right) - \frac{a_n}{\bar{D}}\ln\left(\frac{r}{r_0}\right)\right]. \tag{34b}$$

(Note: The term $g_2$ degenerates in temperature's radial profile, thus we use the higher term $g_4$.)

Hence, the polytropic index is approximated by

$$\nu(r) \cong \tfrac{1}{2}\bar{D}\cdot\left\{1+\left(\frac{\sigma}{\bar{D}}\right)^2 \tfrac{1}{2}\bar{D}\left[\ln\left(\frac{T_0}{T_*}\right) - \frac{2a_n}{\bar{D}}\ln\left(\frac{r}{r_0}\right)\right]\right\}, \tag{35a}$$

$$\gamma(r) \cong 1 + \frac{2}{\bar{D}} - \left(\frac{\sigma}{\bar{D}}\right)^2 \cdot \left[\ln\left(\frac{T_0}{T_*}\right) - \frac{2a_n}{\bar{D}}\ln\left(\frac{r}{r_0}\right)\right]. \tag{35b}$$

The deviation of the polytropic index from its value for ideal adiabatic process is important expression involved in heating formula. This deviation is given by

$$\frac{\gamma_a - \gamma(r)}{\gamma_a - 1} = 1 - \frac{D_a}{\bar{D}}\cdot\left\{\left[1+\left(\frac{\sigma}{\bar{D}}\right)^2 \tfrac{1}{2}\bar{D}\ln\left(\frac{T_0}{T_*}\right)\right]^2 - \left(\frac{\sigma}{\bar{D}}\right)^2 2a_n\ln\left(\frac{r}{r_0}\right)\right\}^{-\tfrac{1}{2}}, \tag{36a}$$

and approximated by

$$\frac{\gamma_a - \gamma(r)}{\gamma_a - 1} \cong 1 - \frac{D_a}{\bar{D}} + \left(\frac{\sigma}{\bar{D}}\right)^2 \cdot \tfrac{1}{2}D_a\left[\ln\left(\frac{T_0}{T_*}\right) - \frac{2a_n}{\bar{D}}\ln\left(\frac{r}{r_0}\right)\right]. \tag{36b}$$

When the nonfluctuating polytropic process is adiabatic, $\gamma_{\sigma=0} = \gamma_a$ or $\bar{D} = D_a$, then,

$$\frac{\gamma_a - \gamma(r)}{\gamma_a - 1} \cong \sigma^2 \cdot \frac{1}{2D_a}\left[\ln\left(\frac{T_0}{T_*}\right) - \frac{2a_n}{D_a}\ln\left(\frac{r}{r_0}\right)\right]. \tag{36c}$$

The deviation of the polytropic index from its value for ideal adiabatic process is proportional to the variance of the fluctuations $\sigma^2$ and restores the adiabatic process when $\sigma$ tends to zero. In general, for any fluctuating polytropic process, the deviation is given by both (i) the deviation of the nonfluctuating indices, and (ii) the part proportional to $\sigma^2$. Consequently, both of these factors contribute to heating.

### 5.3. Heating, dq(r)/dt

We use Eq.(25b) that connects the heating gradient and rate with the difference of polytropic and adiabatic indices,

$$\frac{dq}{dr} = \tfrac{1}{2}a_n\theta_0^2 \cdot \frac{\gamma_a - \gamma(r)}{\gamma_a - 1}\cdot\frac{T(r)}{T_0}\cdot r^{-1}, \tag{37a}$$

$$\frac{dq}{dt} = u_b(r)\cdot\frac{dq}{dr} = u_{b0}r^{-a_u}\cdot\frac{dq}{dr}, \tag{37b}$$

where $a_n k_B T_0 / m = \tfrac{1}{2}a_n\theta_0^2$ and $\theta_0 = \sqrt{2k_B T_0/m}$ denotes the thermal speed. Then, we substitute the radial profiles of the polytropic index and temperature, given respectively in Eq.(32b) and Eq.(32c). Then, we find with the expression of the logarithmic gradient of heating rate,

$$\frac{1}{kT_*/m}\cdot\frac{dq}{d\ln r} = a_n\cdot\left[1 - \frac{D_a}{\bar{D}}\cdot G_\sigma(r)^{-1}\right]\cdot e^{\tfrac{1}{\tfrac{1}{2}\bar{D}}\left(\frac{\sigma}{\bar{D}}\right)^{-2}\cdot[G_\sigma(r)-1]}. \tag{38}$$

Figure 3(b) shows the radial dependence of heating for various values of the variance $\sigma$.

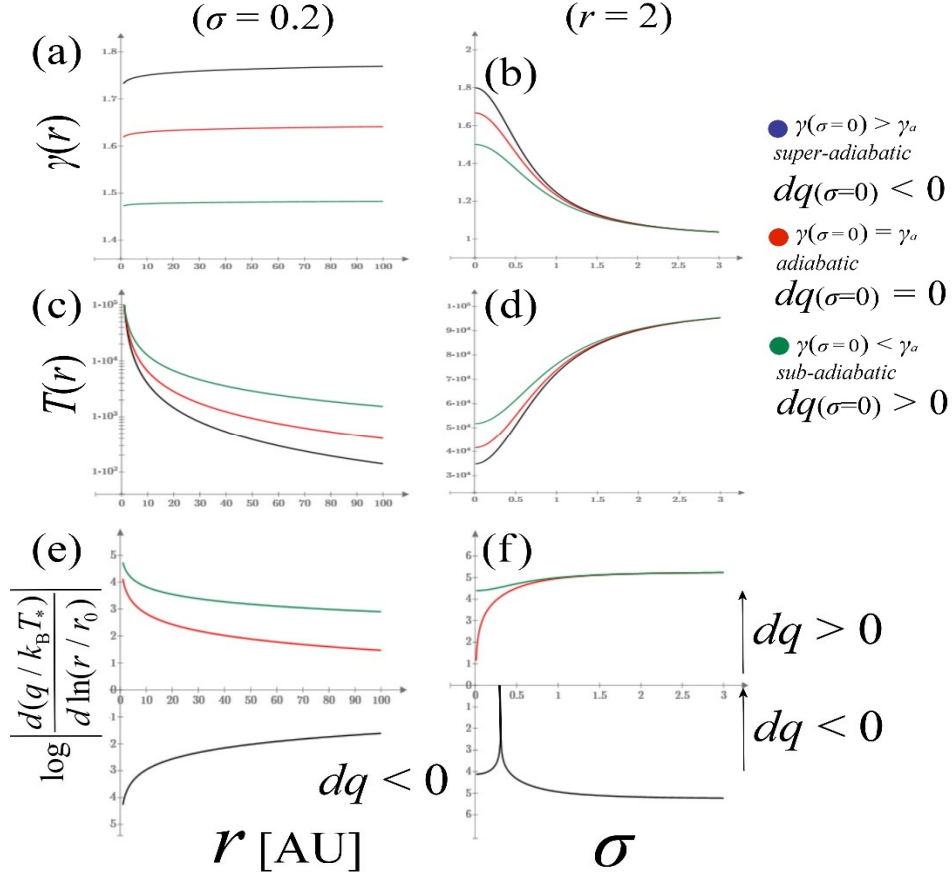

**Figure 3.** Polytropic index $\gamma$ (upper), temperature (middle) and heating (lower) radial profiles and $\sigma$-dependence. Left panels are plotted as a function of heliocentric distance (in AU, for $\sigma$=0.2) and right panels are plotted as a function of the standard deviation $\sigma$ (for $r = 2$ AU), following Eq.(36), and for the three cases of (i) $\gamma(\sigma = 0) > \gamma_a$ or $\bar{D} = 2 < D_a = 3$ (superadiabatic process, in dark blue), (ii) $\gamma(\sigma = 0) = \gamma_a$ or $\bar{D} = D_a = 3$ (adiabatic process, in red), and (iii) $\gamma(\sigma = 0) < \gamma_a$ or $\bar{D} = 4 > D_a = 3$ (subadiabatic process, in green). (Note: heating is plotted on a double log scale, using $sign(dq) \cdot \log|dq|$, to distinguish the cases of heat entering ($dq$>0) and exiting ($dq$<0) the system.

Next, we derive the expression of heating for small sigma, which can be written as

$$\frac{dq}{dr} \cong A \cdot \left(\frac{r}{r_0}\right)^{-B} \cdot \left[1 - C \cdot \ln\left(\frac{r}{r_0}\right)\right], \tag{39a}$$

where

$$A = \left[1 - \frac{D_a}{\bar{D}} + \tfrac{1}{2} D_a \ln\left(\frac{T_0}{T_*}\right)\left(\frac{\sigma}{\bar{D}}\right)^2\right] a_n, \quad B = 1 + \frac{a_n}{\tfrac{1}{2}\bar{D}},$$

$$C = \frac{\left[\left(1 - \frac{D_a}{\bar{D}}\right)\ln\left(\frac{T_0}{T_*}\right) - \frac{D_a}{\bar{D}}\right] a_n}{1 - \frac{D_a}{\bar{D}} + \tfrac{1}{2} D_a \ln\left(\frac{T_0}{T_*}\right)\left(\frac{\sigma}{\bar{D}}\right)^2} \cdot \left(\frac{\sigma}{\bar{D}}\right)^2. \tag{39b}$$

As expected, heating becomes zero for nonfluctuating adiabatic processes, while polytropic processes, in general, have two components contributing to heating, the nonadiabatic polytropic index, $\bar{D} \neq D_a$, and the presence of fluctuations with non-zero variance $\sigma^2$.

## 6. Connection of fluctuating polytropic processes with heliospheric turbulence

### 6.1. Rationale

We observe that the logarithmic part of the heating expression in Eq.(39a) does not exist for nonfluctuating processes ($\sigma = 0$, i.e., $C = 0$). We analyze the expression in two components; the first component is equivalent to a nonfluctuating process (and is independent of $\sigma$), while the second component corresponds to the fluctuations (and is proportional to $\sigma^2$), i.e.,

$$\frac{dq_{all}}{dr} = \left(\frac{dq}{dr}\right)_{\sigma=0} + \left(\frac{dq}{dr}\right)_\sigma , \qquad (40a)$$

where the nonfluctuating term is

$$\frac{1}{kT_0/m} \cdot \left(\frac{dq}{d\ln r}\right)_{\sigma=0} = \left(1 - \frac{D_a}{\bar{D}}\right) a_n \cdot \left(\frac{r}{r_0}\right)^{-\frac{2a_n}{\bar{D}}} , \qquad (40b)$$

and the fluctuating term is proportional to $\sigma^2$ and equal to

$$\frac{1}{kT_0/m} \cdot \left(\frac{dq}{d\ln r}\right)_\sigma = \left(\frac{\sigma}{\bar{D}}\right)^2 \cdot \tfrac{1}{2} D_a \ln\left(\frac{T_0}{T_*}\right) a_n \cdot \left(\frac{r}{r_0}\right)^{-\frac{2a_n}{\bar{D}}} \cdot \left\{ 1 + \frac{\left[\left(1 - \frac{D_a}{\bar{D}}\right)\ln\left(\frac{T_0}{T_*}\right) - \frac{D_a}{\bar{D}}\right]}{\tfrac{1}{2} D_a \ln\left(\frac{T_0}{T_*}\right)} a_n \cdot \ln\left(\frac{r}{r_0}\right) \right\} , \qquad (40c)$$

which is again written in the form of Eq.(39a), with $A$, $B$, and $C$, now given by

$$A = \left(\frac{\sigma}{\bar{D}}\right)^2 \cdot \tfrac{1}{2} D_a \ln\left(\frac{T_0}{T_*}\right) a_n , \quad B = 1 + \frac{a_n}{\tfrac{1}{2}\bar{D}} , \quad C = \frac{\left[\left(1 - \frac{D_a}{\bar{D}}\right)\ln\left(\frac{T_0}{T_*}\right) - \frac{D_a}{\bar{D}}\right] a_n}{\tfrac{1}{2} D_a \ln\left(\frac{T_0}{T_*}\right)} . \qquad (41)$$

Therefore, while the nonfluctuating heating is expressed by a power-law, the fluctuating heating also includes the logarithmic term proportional to the parameter $C$.

The difference in the expressions of the two components is important to understand their physical interpretation and origin. While the first heating component can be the result of any coherent physical source, the second component can be the result of turbulence. In order to understand this assertion, we first need to examine the radial profile of turbulent heating.

### 6.2. Radial profile of turbulent heating

The turbulent heating can be modelled by a set of 3D differential equations (e.g., Smith et al. 2001; 2006) that involve the transport of turbulent energy and the energy cascade at intermediate scales. For a radially symmetric flow, the model provides steady-state equations for the radial evolution of the turbulent energy $Z^2 = E_t$ (averaged between the forward $E_t^+$ and backward $E_t^-$ propagating modes), the turbulence correlation length $\lambda$, and the proton temperature $T$:

$$\frac{d}{dr}Z^2 = (C_{sh} - M\sigma_D - 1)\cdot\frac{Z^2}{r} - \alpha\frac{Z^3}{\lambda u_b}, \quad (42a)$$

$$\frac{d}{dr}\lambda = (M\sigma_D - \hat{C})\cdot\frac{\lambda}{r} + \frac{\beta}{u_b}\cdot Z, \quad (42b)$$

$$\frac{dT}{dr} = -a_n(\gamma_a - 1)\cdot\frac{T}{r} + \frac{m_p}{k_B}\cdot(\gamma_a - 1)\cdot\frac{dE_t}{dr}. \quad (42c)$$

The involved parameters are the following: mixing constant $M=1/3$, normalized energy difference $\sigma_D=-1/3$, turbulent viscosity terms $\alpha=0.8$ and $\beta=\alpha/2=0.4$ (where the choice of $\alpha/\beta=2$ preserves the turbulent viscosity, W. H., Matthaeus et al. 1996), constants related to shear driving turbulent energy $C_{sh}=1.4$ and $\hat{C}$ (with no clearly set value), and the PUI driven turbulent energy $E_{t,pui}$. (The bulk speed $u_b$ represents the solar wind speed.) The first two equations constitute a 2D system of differential equations between $Z^2$ and $\lambda$. The third equation gives the transport equation of temperature (e.g., see G.P. Zank et al. 2021; Livadiotis & McComas 2023a), where the gradient of the turbulent energy has been substituted by

$$\frac{dE_t}{dr} = \tfrac{1}{2}\alpha\cdot\frac{Z^3}{u_b\lambda}. \quad (43)$$

In Appendix B, we solve this system of differential equations to derive the turbulent energy $Z^2(r)$ and the correlation length $\lambda(r)$, with respect to the heliospheric radius $r$ in the inner heliosphere. The subsystem of the first two equations is solved to derive $Z^2(r)$ and $\lambda(r)$, and then, we substitute in Eq.(43) to derive the gradient of the turbulent energy (e.g., see L. Adhikari et al. 2015; G. Livadiotis 2019b; G. Livadiotis et al. 2020), that is,

$$\frac{dE_t}{dr} \cong \frac{\tfrac{1}{2}\alpha Z_0^3}{\lambda_0 u_{b0}}\cdot\left(\frac{r}{r_0}\right)^{-(2+M\sigma_D-C_{sh})}\cdot\left[1 - \frac{2\alpha Z_0 r_0}{\lambda_0 u_{b0}}\cdot\ln\left(\frac{r}{r_0}\right)\right], \quad (44)$$

which is identical to the heating of a fluctuating process, shown in Eq.(40c), i.e.,

$$\left(\frac{dq}{dr}\right)_\sigma = \frac{a_n D_a \sigma^2 kT_0}{2\overline{D}^2 mr_0}\cdot\ln\left(\frac{T_0}{T_*}\right)\cdot\left(\frac{r}{r_0}\right)^{-\frac{2a_n}{\overline{D}}-1}\cdot\left[1 - \frac{\frac{D_a}{\overline{D}} - \left(1-\frac{D_a}{\overline{D}}\right)\ln\left(\frac{T_0}{T_*}\right)}{\frac{1}{2a_n}D_a\ln\left(\frac{T_0}{T_*}\right)}\cdot\ln\left(\frac{r}{r_0}\right)\right]. \quad (45)$$

We observe that turbulent heating has the same expression as the heating of fluctuating polytropic processes, and thus can be expressed in the unique form

$$\frac{dq_\sigma}{dr} = \frac{dE_t}{dr}$$

$$\cong A \cdot \left(\frac{r}{r_0}\right)^{-B} \cdot \left[1 - C \cdot \ln\left(\frac{r}{r_0}\right)\right], \quad (46a)$$

where the involved parameters can be expressed either in terms of thermodynamics or turbulence,

$$A = \frac{a_n D_a \sigma^2 k T_0}{2\bar{\bar{D}}^2 m r_0} \cdot \ln\left(\frac{T_0}{T_*}\right) = \frac{\tfrac{1}{2}\alpha Z_0^3}{\lambda_0 u_{b0}},$$

$$B = \frac{2a_n}{\bar{\bar{D}}} + 1 = 2 + M\sigma_D - C_{sh} - \varepsilon, \quad (46b)$$

$$C = \frac{\frac{D_a}{\bar{\bar{D}}} - \left(1 - \frac{D_a}{\bar{\bar{D}}}\right)\ln\left(\frac{T_0}{T_*}\right)}{\frac{1}{2a_n} D_a \ln\left(\frac{T_0}{T_*}\right)} = \frac{(\tfrac{3}{2}\alpha + \beta)Z_0 r_0}{\lambda_0 u_{b0}}.$$

(The arbitrary constant $\varepsilon$ is determined by Eqs.(B10,B17) in Appendix B.)

### *6.3. Interpretation of heating associated with fluctuating polytropic processes*

The heating of nonfluctuating polytropic process is only connected to deviation of the polytropic index and its deviation from the adiabatic value, where heating is entering the system, $dq > 0$, when $\gamma < \gamma_a$ (or $\bar{D} > D_a$), while heating is exiting the system, $dq < 0$, when $\gamma > \gamma_a$ (or $\bar{D} < D_a$); trivially, no heating is associated with nonfluctuating adiabatic process. On the other hand, fluctuating polytropic process have an additional component of heating entering the system, solely caused by the fluctuations. The two components have complementary contributions to the total heating, $dq_{tot} = dq_{\sigma=0} + dq_\sigma$.

The turbulent heating, $dE_t$ has the same radial profile as the heating associated with the fluctuating polytropic process, $dq_\sigma$, while it differs from the component associated with the deviation of polytropic index from its adiabatic value, $dq_{\sigma=0}$. In particular, the radial profile of $dq_{\sigma=0}$ is a power-law, while the profile of $dq_\sigma$ has the additional logarithmic dependence, which makes the whole expression identical to that of turbulent heating, $dq_\sigma = dE_t$.

Therefore, turbulence does not affect thermodynamics by decreasing the polytropic index. This is quite a surprising result! Previous related analyses sought to link heating to be related to the deviation between the local polytropic $\gamma = 1 + 2/\bar{D}$ and the adiabatic index $\gamma_a = 1 + 2/D_a$ (=5/3 for $D_a$=3) as a way to identify and estimate the turbulent heating (e.g., M.K. Verma et al. 1995; B.J. Vasquez et al. 2007; L. Sorriso-Valvo et al. 2007; R. Marino et al. 2008; G. Livadiotis 2015; 2019a; 2021; E. Gravanis et al. 2019; G. Livadiotis et al. 2020; G. Livadiotis & D.J. McComas 2023a). Here we show that this cannot be true, because any type of nonturbulent heating can be connected with this polytropic index deviation, but the turbulent heating can be only connected through the fluctuations of polytropic indices. Specifically, the turbulent heating has a specific mathematical form that cannot be matched with the expression of heating due to the deviation of the polytropic index from the corresponding adiabatic index; instead, it matches the expression of heating originating from fluctuations of the polytropic indices.

Therefore, understanding fluctuating polytropic processes reveals that turbulent and nonturbulent heating have separate, distinguishable thermodynamics. Turbulent heating occurs when energy cascades from large-scale motions to small scales, where it dissipates. Nonturbulent heating does not rely on turbulent cascades or stochastic fluctuations. It is associated with the formation of organized patterns or structures (e.g., phase-correlated electromagnetic fields or waves), shock compression, and collisional heating. (See: B. T. Tsurutani et al. 2023).

### *6.4. Radial profile of PUI turbulent energy*

The main source of heating the solar wind plasma through its expansion in the heliosphere is from energy extracted for the solar wind bulk flow via PUIs. The energy transfer from PUIs is an example of both turbulent and nonturbulent heating sources. C.W. Smith et al. (2006) studied specifically these two components of PUI heating providing the radial profile of the turbulent heating and of the total, turbulent and nonturbulent heating. In particular, these authors clearly describe that PUI energy transfer that heats the solar wind protons "… *contributes to turbulent fluctuations as well and is not exactly equivalent to the plasma heating rate due to pickup protons.*"

The thermodynamic description of these two components is the aforementioned (i) deviation of polytropic and adiabatic indices, and (ii) the fluctuations of the polytropic index. Through this correspondence, we can derive the ratio of the turbulent over the total heating, and its complementary, the ratio of the nonturbulent over the total heating; there are, respectively, given by

$$\frac{dE_t}{dq_{all}} = \frac{dE_t}{dE_t + dE_{non-t}} = \frac{\frac{1}{2}D_a \ln\left(\frac{T_0}{T_*}\right)\left(\frac{\sigma}{\overline{D}}\right)^2 \cdot \left\{1 - \frac{\frac{D_a}{\overline{D}} - \left(1 - \frac{D_a}{\overline{D}}\right)\ln\left(\frac{T_0}{T_*}\right)}{\frac{1}{2}D_a \ln\left(\frac{T_0}{T_*}\right)} a_n \ln\left(\frac{r}{r_0}\right)\right\}}{1 - \frac{D_a}{\overline{D}} + \frac{1}{2}D_a \ln\left(\frac{T_0}{T_*}\right)\left(\frac{\sigma}{\overline{D}}\right)^2 \cdot \left\{1 - \frac{\frac{D_a}{\overline{D}} - \left(1 - \frac{D_a}{\overline{D}}\right)\ln\left(\frac{T_0}{T_*}\right)}{\frac{1}{2}D_a \ln\left(\frac{T_0}{T_*}\right)} a_n \ln\left(\frac{r}{r_0}\right)\right\}}, \quad (47a)$$

and

$$\frac{dE_{non-t}}{dq_{all}} = \frac{dE_{non-t}}{dE_t + dE_{non-t}} = \frac{1 - \frac{D_a}{\overline{D}}}{1 - \frac{D_a}{\overline{D}} + \frac{1}{2}D_a \ln\left(\frac{T_0}{T_*}\right)\left(\frac{\sigma}{\overline{D}}\right)^2 \cdot \left\{1 - \frac{\frac{D_a}{\overline{D}} - \left(1 - \frac{D_a}{\overline{D}}\right)\ln\left(\frac{T_0}{T_*}\right)}{\frac{1}{2}D_a \ln\left(\frac{T_0}{T_*}\right)} a_n \ln\left(\frac{r}{r_0}\right)\right\}}. \quad (47b)$$

Furthermore, the ratio $dE_t / dq_{all}$ can be used to estimate the turbulent heating of PUIs. For this, we need to first derive the radial profile of the total heating, $dq_{all} / dr$, and then, we use the chain equation $dE_t / dr = (dE_t / dq_{all}) \cdot (dq_{all} / dr)$. For deriving $dq_{all} / dr$, we use the radial profile of the polytropic index $\gamma(r) = 5/3 - 0.0051 r^{1.7} / (1 + 0.0034 r^{1.7})$ (G. Livadiotis & D.J. McComas 2023a). This is applied into the thermodynamic equation that connects it to the total heating, Eq.(25),

$$\frac{dq_{all}}{dr} = \frac{a_n k_B T_0}{m r_0} \cdot \frac{\gamma_a - \gamma(r)}{\gamma_a - 1} \cdot \frac{T(r)}{T_0} \cdot \left(\frac{r}{r_0}\right)^{-1}, \quad (48)$$

where we expressed the temperature profile in terms of the polytropic index $\gamma(r)$, namely, $T(r)/T_0 = \exp\{-a_n \int_{r_0}^{r} [\gamma(r) - 1] r^{-1} dr\}$; alternatively, we use the radial profiles from D.J. McComas (2025a). Then, we estimate the rate of turbulent heating through

$$\frac{dE_{tPUI}}{dt} = \left(\frac{dE_t}{dq_{all}}\right) \cdot \left(\frac{dq_{all}}{dr}\right) \cdot u_b(r). \quad (49)$$

Figure 4(a) plots the PUI turbulent heating as derived from Voyager 2 observations by C.W. Smith et al. (2006), for which we fit the modelled radial profile of the PUI turbulent heating rate as given in Eq.(49), after substituting Eq.(47a) and Eq.(48). As shown in Figure 4(b), the deviations of the observed rates above and below the modelled ones are correlated with solar activity maxima and minima. Therefore, in order to better describe the model, we adopt two components of polytropic fluctuations, (i) one component $\sigma_0$ that exists independently of solar activity and remains stationary along the evolution of PUI distribution throughout the heliosphere, and (ii) a second component $\sigma_{SN}$ that evolves proportional to the sunspot number (normalized accordingly to its maximum value). Overall, we have

$$\sigma^2_{total} = \sigma^2_0 + \sigma^2_{SN}, \text{ with } \sigma_{SN} = \zeta \cdot \sigma_0 \cdot \frac{SunspotNumber}{half \ of \ SunspotNumber_{max}}, \quad (50)$$

The local effective degrees of freedom are also scaled with $\sigma_0$, as

$$\bar{D} = D_a + A_D \cdot (\sigma^2_0)^{a_D}. \quad (51)$$

Then, the model describes better the observations of turbulent heating rates, as shown in Figure 4(c). The fit of the model estimates the involved parameters values, i.e., $\zeta = 2$, $A_D \cong 570$ and $a_D \cong 1.17$, but it is independent of the value of $\sigma_0$ as it always has small value. Here we use $\sigma_0 = 0.01$, thus $\bar{D} = 3.012$ (for $D_a = 3$). (For the optimization of the model to observations, we have also used the thermal scale $T_* = 1\,\text{K}$ and the initial values $T_0 = 3 \cdot 10^4\,\text{K}$, $u_{b0} = 400\,\text{km s}^{-1}$.)

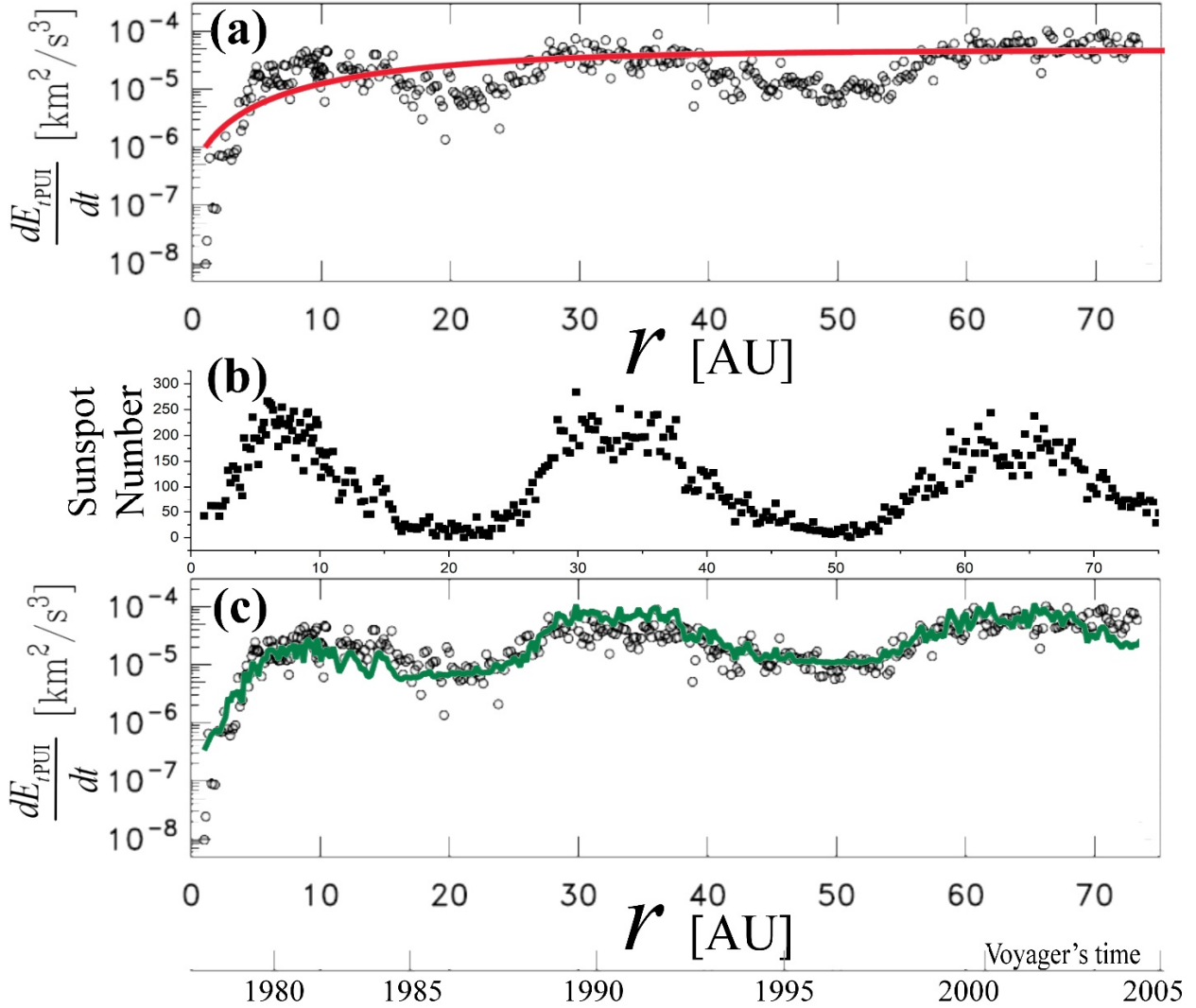

**Figure 4.** PUI turbulent heating modelled thermodynamically through a fluctuating polytropic process. (a) Fit of PUI turbulent heating rate to observations for a fixed variance of fluctuations, $\sigma^2_0$. (b) The sunspots number is correlated with the deviations of the observed from the modelled rates. (c) Fit of PUI turbulent heating rate to observations for a variance given by a fixed component $\sigma^2_0$ and a nonstationary component $\sigma^2_{SN}$, where $\sigma_{SN}$ is proportional to sunspot number.

## 7. Discussion and conclusions

In this paper we studied the heating associated with fluctuating polytropic processes. The first question we faced was to determine whether fluctuating adiabatic processes produce a nonzero overall heating of the flowing particle system. We found that heating becomes zero for nonfluctuating adiabatic processes, as expected, but there is a nonzero positive value of heating entering the flowing system for fluctuating adiabatic processes. In general, there are two components contributing to the heating of fluctuating polytropic processes, (i) the nonadiabatic polytropic index, $\gamma \neq \gamma_a$ or $\bar{D} \neq D_a$, and (ii) the presence of fluctuations with variance $\sigma^2$. The first component contributes either with heat entering the system for the subadiabatic processes, $\gamma < \gamma_a$ or with heat exiting the system for superadiabatic processes, $\gamma > \gamma_a$. The second component that stands for fluctuating polytropic processes, which always contribute as heating entering the system. Furthermore, we derived the analytical expressions of radial profiles of the heating rates for fluctuating polytropic processes, which have been distinguished by the two components of adiabatic deviation and of polytropic fluctuations. In order to interpret the nature of the two components, we revisited the analytical modeling of turbulent heating and found that the radial profiles of the heating rate associated with polytropic fluctuations and of the turbulence are surprisingly identical. This suggests that turbulence, unavoidably, heats plasma particle populations through fluctuations of polytropic processes and not through polytropic indices that deviating from the adiabatic expansion.

First, we developed a general way of expressing a multi-polytrope, as a superposition of polytropes. While a polytrope is described by a power-law relationship among thermodynamic parameters (Figure 1), typically, between density and temperature, the multi-polytrope can be described by any analytical function expressed as a discrete (sum) or continuous (integral) superposition of powers of temperature (Figure 2). The summation or integration holds for the effective degrees of freedom, that is, a fractional number of dimensionalities, as the velocity distribution does not fully populate the available unitary-dimensionality phase space. Then, we applied the superposition in the particular case where the nonconstant polytropic index is constructed by random fluctuations of polytropic processes. The distribution of the effective degrees of freedom is given by a normal distribution of random fluctuations. We also found the same results for any distribution function of random fluctuations but with small variance. Then, we derived the generalized polytropic relationship, temperature radial profile, and heating rates that characterized fluctuating polytropic processes; (we expressed heating using rate, gradient, and heat capacity).

We derived the expression of the polytropic index and heating of particle systems flowing under fluctuating polytropic processes. In the case of fluctuating adiabatic processes, heating is proportional to the variance of the fluctuations, while in the general case of fluctuating polytropic processes, heating is entering the system because of two thermodynamic factors: (i) the difference between the nonfluctuating (ideal) polytropic and the ideal adiabatic indices or degrees of freedom, and (ii) the nonzero variance that measures the fluctuations of the polytropic process.

In the case of fluctuating adiabatic processes, the polytropic index was found to be smaller than the respective ideal adiabatic, thus heat enters the system; the fluctuation of polytropic process has an impact similar to that of subadiabatic processes, for which heating is entering the system, as opposed to super-adiabatic processes, for which heat is leaving the system. The effect of fluctuations in heating is summarized as follows: The temperature radial profile was found to be decreasing with distance less than the decrease of the adiabatic process (adiabatic cooling) by a factor that is proportional to the variance of the normally distributed fluctuations. This is caused by the heating entering the system due to the fluctuations. The corresponding polytropic index is smaller than the adiabatic, decreasing with increasing the logarithm of radial distance. Recall that the polytropic index decreases while heating increases with increasing variance of the polytropic fluctuations (Figure 3).

Finally, we showed that the analytical radial profiles of heating of fluctuating polytropic processes and of turbulent heating are identical. In particular, we derived the analytical solution of a differential system that models the turbulent heating, and then, compared it with the heating of fluctuating polytropic processes. We found that the two expressions of heating are identical. This suggests that turbulence heats plasma particle populations by fluctuating their polytropic processes and not by deviations of the polytropic index, as was previously thought (e.g. G. Livadiotis & D.J. McComas 2023a; 2025a). The thermodynamics of polytropic processes unfolds a connection with two different types of heating, the turbulent and nonturbulent heating. The nonturbulent heating is connected with the deviation of polytropic index from its adiabatic value, while the turbulent heating is connected with the fluctuations of the polytropic index. Moreover, the percentage of the turbulent heating over the overall turbulent and nonturbulent heating can be solely described through thermodynamics, that is, the local values of polytropic index, $\gamma = 1 + 2/\bar{D}$, and the variance of the polytropic fluctuations, $\sigma^2$. We have applied the theory to estimate the turbulent heating rates of PUIs through their (nonadiabatic) expansion in the heliosphere. The derived radial profile fits very well the observation turbulent heating rates. We found that the variance of fluctuations is the sum of two components, (i) a stationary component that remains fixed along the evolution of PUI distribution throughout the heliosphere, and (ii) a second component that varies proportionally to the sunspot number.

In general, given a radial profile of the polytropic index $\gamma(r)$, we can estimate the average degrees of freedom $\bar{D}$ and its variance $\sigma^2$ by fitting the respective analytical equation of polytropic index, Eq.(15b), to the given profile. Alternatively, we can analytically derive the temperature profile from $T(r) = T_0 \cdot \exp\{-a_n \int_{r_0}^{r} [\gamma(r)-1] r^{-1} dr\}$, while the variance can be determined from more detailed analytical forms, such as, $\sigma^2 = 2\bar{D} \cdot d\nu(r)/d\ln T(r)$ or $\sigma^2 = -(\bar{D}^3 / 2a_n^2) \cdot d^2 \ln T(r)/d(\ln r)^2$. Then, we can calculate the total heating associated with the fluctuating polytropic process, $dq_{all}/dt$, and its portion that corresponds to the nonturbulent $(dq/dt)_{\sigma=0}$ and turbulent $(dq/dt)_\sigma$ heating components. (See Figure 5).

We remark that the treatment of determining the turbulent (and non-turbulent) heating remains unchanged even when the pressure tensor is anisotropic. In such cases, the plasma exhibits different temperature components along each physical dimension (kinetic degrees of freedom); e.g., $T_\parallel$ in the direction parallel to the magnetic field, and $T_\perp$ for the two perpendicular directions, with the anisotropy commonly described by the ratio $\alpha = T_\perp / T_\parallel$. The thermal components are not true "temperatures", because they do not conform to thermodynamics. The actual temperature $T$ is given by their average, e.g., $T = (T_\parallel + 2T_\perp)/3$, and this is the temperature that enters the polytropic description of plasma processes. The anisotropy is known to affect the adiabatic degrees of freedom $D_a$ (and thereby, the adiabatic index), but does not influence the fluctuations of the effective degrees of freedom (G. Livadiotis & G. Nicolaou 2021; C. Katsavrias et al. 2024).

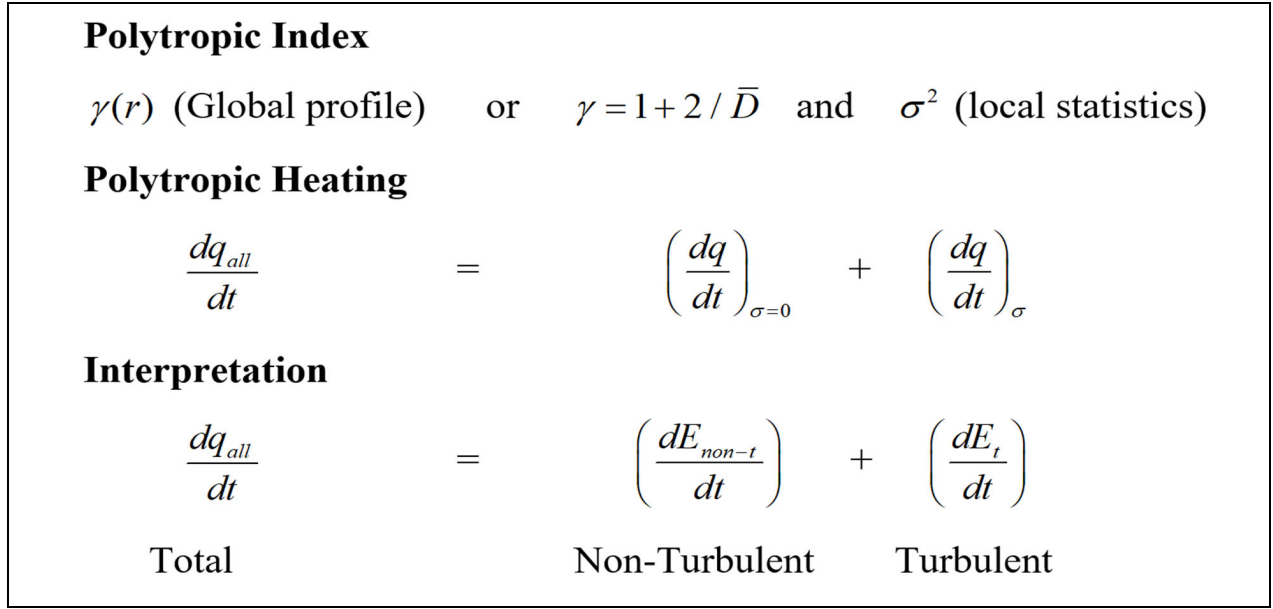

**Figure 5.** Polytropic index variability can be connected with the nature of heating. Radial dependence of polytropic index is connected with the total heating rate, while its analysis to an average polytropic index fluctuating with a variance (local measurements) leads to further interpretation, as they are connected with the nonturbulent and turbulent heating rates, respectively.

The developments of this paper are ready to be used in further theoretical analyses and/or applications. Open topics for investigation and/or application are the following: (i) Detection of parabolic behavior in space plasmas, can be linked to fluctuating adiabatic or other polytropic processes. Identifying the value of polytropic index and variance can be used to estimate the turbulent heating. Such parabolic polytropic relationships have already been observed, e.g., in the inner heliosphere using measurements from Parker Solar Probe (Fox et al. 2016; Nicolaou et al. 2020) and the heliosheath using measurements from Interstellar Boundary Explorer (D.J. McComas et al. 2009a;b; G. Livadiotis & D.J. McComas 2013; G. Livadiotis 2016). (ii) Local measurements of polytropic indices can also lead to an average polytropic index and its variance. Comparing the local measurements with the global radial profile of polytropic index, we can verify whether the superposition refers to the fluctuations of polytropic processes or to more complicate analysis. (iii) Once the average polytropic index and its variance are measured, then, the heating rate transferred into the system and its components to turbulent and nonturbulent heating, can be fully estimated. This is in contrast to previous analyses where there was no mention to nonturbulent heating. (iv) Theoretical investigation of random fluctuations described by more general statistics, such as, the *q*-statistics (e.g., see: Livadiotis 2017; C. Tsallis 2023; L.Y. Khoo et al. 2025); these can be linked to more general expressions of turbulent heating. Heliospheric observations from the Interstellar Mapping and Acceleration Probe (IMAP) mission (McComas et al. 2018; 2025**b**) will allow for more detailed characterization of fluctuating polytropic processes, and the heating of solar wind protons from PUIs (also, see: J. S. Rankin et al. 2025; S. A. Livi et al. 2025; T. S. Horbury et al. 2025; E. R. Christian et al. 2025; D.B. Reisenfeld 2025).

**Acknowledgments**
This work was funded in part by the IMAP mission as a part of NASA's Solar Terrestrial Probes (STP) Program (grant No. 80GSFC19C0027).

**ORCID iDs**
G. Livadiotis https://orcid.org/0000-0002-7655-6019
D. J. McComas https://orcid.org/0000-0001-6160-1158

**Appendix A. Heat Capacity of polytropic processes**

In this Appendix, we derive the overall heat exchange of solar wind protons, which are subject to fluctuations of a polytropic process. This is carried out by calculating the heating rate and gradient, as well as the heat capacity of this plasma. To achieve this, we develop a general way of constructing the nonconstant polytropic index from a multi-polytropic process.

A convenient way of expressing the heating in systems with varying polytropic processes is through the heat capacity. It defines how much a particle system must be heated for its temperature to be raised by 1 deg K and is useful for understanding the thermodynamics of the system, as well as its interactions with other systems. The heat capacity depends on the thermodynamic process that the system undergoes, when it experiences a change in its state variables (e.g., pressure, volume, temperature, internal energy, entropy), due to interactions with its environment. First, we derived the heat capacity for an ideal polytropic process, which is characterized by a single polytropic index (i.e., a polytrope), and then, we proceed to the case of fluctuating polytropic processes.

Starting from Eq.(23), and given the identity $T \cdot d \ln V = -dT \cdot d \ln n / d \ln T$, we derive the heat capacity that characterizes polytropic processes with index $\gamma$,

$$C_\gamma / k_B \equiv \frac{dq}{k_B dT} = \frac{\gamma - \gamma_a}{(\gamma_a - 1)(\gamma - 1)} = \frac{1}{\gamma_a - 1} - \frac{1}{\gamma - 1} . \quad (A1a)$$

The polytropic $D$ and kinetic $D_a$ degrees of freedom can be similarly defined by $\gamma = 1 + 1/\nu = 1 + 2/D$ and $\gamma_a = 1 + 1/\nu_a = 1 + 2/D_a$; then, Eq.(4) is written in terms of degrees of freedom instead of polytropic indices, thus

$$C_\gamma / k_B = \frac{1}{\gamma_a - 1} - \frac{1}{\gamma - 1} = \nu_a - \nu = \tfrac{1}{2}(D_K - D) . \quad (A1b)$$

Two characteristic processes are isobaric (i.e., constant pressure, $\gamma \to 0$) and isochoric (constant volume, $\gamma \to \infty$) with heat capacities, $C_P = (\tfrac{1}{2}D_a + 1) \cdot k_B$ and $C_V = \tfrac{1}{2}D_a \cdot k_B$, respectively. Their ratio is frequently used as it is equal to the adiabatic index, $C_P / C_V = 1 + 2/D_a = \gamma_a$. Other processes have rather trivial values of heat capacities; two examples are isothermal processes ($\gamma \to 1$) with infinite heat capacity, $C_1 \to \pm\infty$, meaning the system can exchange any amount heat before changing temperature, and adiabatic processes ($\gamma = \gamma_a$) where no heat is exchanged, $C_a = 0$.

Next, we derive the heat capacity of the fluctuating polytropic processes. For this, we substitute the polytropic index, expressed as a function of temperature in Eq.(19a), into the heat capacity equation of Eq.(2b), leading to

$$C_\gamma / k_B = \nu_a - \nu_{\sigma=0} - \sigma^2 \cdot \tfrac{1}{4} \ln(T / T_*) , \quad (A2a)$$

where $\nu_{\sigma=0} = \tfrac{1}{2}\bar{D}$ indicates, again, the polytropic index of the nonfluctuating process.

In the case of a fluctuating adiabatic process, $\nu_{\sigma=0} = \nu_a$, the heat capacity is simplified to

$$C_\gamma / k_B = -\sigma^2 \cdot \tfrac{1}{4} \ln(T / T_*) , \quad (A2b)$$

which is proportional to the variance of the fluctuations of adiabatic process, thus, the stronger the fluctuations, the larger the heat capacity, while it becomes zero in the case of in ideal adiabatic process.

Furthermore, substituting the radial profile of temperature, given in Eq.(A1a), into the expression of heat capacity in terms of temperature, Eq.(28b), we end up with the radial profile of heat capacity, that is,

$$C_\gamma / k_B = \tfrac{1}{2} D_a - \tfrac{1}{2}\overline{D} \cdot G_\sigma(r) \cong \tfrac{1}{2} D_a - \tfrac{1}{2}\overline{D} - \left[ \tfrac{1}{2}\ln\left(\frac{T_0}{T_*}\right) - \frac{a_n}{\overline{D}}\ln\left(\frac{r}{r_0}\right)\right] \cdot \sigma^2. \quad (A3a)$$

In the case of a fluctuating adiabatic process, $\overline{D} = D_a$, the heat capacity is reduced to

$$C_\gamma / k_B \cong -\left[ \tfrac{1}{2}\ln\left(\frac{T_0}{T_*}\right) - \frac{a_n}{D_a}\ln\left(\frac{r}{r_0}\right)\right] \cdot \sigma^2. \quad (A3b)$$

As expected, the heat capacity is zero for the nonfluctuating adiabatic process. However, this is nonzero and proportional to the variance of the polytropic fluctuations, $\sigma^2$, in the case of the fluctuating adiabatic process.

**Appendix B. Model of turbulent energy**

We work with the first two differential equations of the 3D model in Section 6.2, Eqs.(42a,b),

$$\frac{d}{dr} Z^2 = (C_{sh} - M\sigma_D - 1) \cdot \frac{Z^2}{r} - \alpha \frac{Z^3}{\lambda u_b}, \quad (B1a)$$

$$\frac{d}{dr} \lambda = (M\sigma_D - \hat{C}) \cdot \frac{\lambda}{r} + \frac{\beta}{u_b} \cdot Z, \quad (B1b)$$

written as

$$\alpha^{-1}\frac{d}{dr}\ln Z^2 = \alpha^{-1}(C_{sh} - M\sigma_D - 1) \cdot \frac{1}{r} - \frac{Z}{\lambda u_b}, \quad (B2a)$$

$$\beta^{-1}\frac{d}{dr}\ln\lambda = \beta^{-1}(M\sigma_D - \hat{C}) \cdot \frac{1}{r} + \frac{Z}{\lambda u_b}, \quad (B2b)$$

in order to be added to cancel out the term $Z/(\lambda u_b)$, i.e.,

$$\frac{d}{d\ln r}\left(\alpha^{-1}\ln Z^2 + \beta^{-1}\ln\lambda\right) = \alpha^{-1}(C_{sh} - M\sigma_D - 1) + \beta^{-1}(M\sigma_D - \hat{C}), \quad (B3)$$

or

$$\frac{d}{d\ln r}\ln(Z^{2/\alpha}\lambda^{1/\beta}) = \alpha^{-1}(C_{sh} - M\sigma_D - 1) + \beta^{-1}(M\sigma_D - \hat{C}) = const., \quad (B4)$$

Since the right-hand side term is constant, we obtain

$$Z^{2/\alpha}\lambda^{1/\beta} \propto r^{\alpha^{-1}(C_{sh} - M\sigma_D - 1) + \beta^{-1}(M\sigma_D - \hat{C})}, \quad (B5)$$

or

$$Z = Z_0 \lambda_0^{\tfrac{1}{2}\alpha/\beta} \lambda^{-\tfrac{1}{2}\alpha/\beta} (r/r_0)^{\tfrac{1}{2}(C_{sh} - M\sigma_D - 1) + (\tfrac{1}{2}\alpha/\beta)(M\sigma_D - \hat{C})}, \quad (B6a)$$

$$\lambda = \lambda_0 Z_0^{\beta/\tfrac{1}{2}\alpha} Z^{-\beta/\tfrac{1}{2}\alpha} (r/r_0)^{(\beta/\alpha)(C_{sh} - M\sigma_D - 1) + (M\sigma_D - \hat{C})}. \quad (B6b)$$

Substituting (B6b) back to Eq.(B1a), we obtain

$$\frac{d}{dr}Z^2 = (C_{\text{sh}} - M\sigma_{\text{D}} - 1)\cdot\frac{Z^2}{r} - \alpha\frac{Z_0^{-\beta/\frac{1}{2}\alpha}}{u_{b0}\lambda_0}Z^{3+\beta/\frac{1}{2}\alpha}(r/r_0)^{-(\beta/\alpha)(C_{\text{sh}}-M\sigma_{\text{D}}-1)-(M\sigma_{\text{D}}-\hat{C})+a_u}, \tag{B7}$$

After some calculus, we derive the 1D differential equation of $Z^{-1-\beta/\frac{1}{2}\alpha}$, i.e.,

$$\frac{d}{dr}Z^{-1-\beta/\frac{1}{2}\alpha} + \frac{(1+\beta/\frac{1}{2}\alpha)\frac{1}{2}(C_{\text{sh}}-M\sigma_{\text{D}}-1)}{r}\cdot Z^{-1-\beta/\frac{1}{2}\alpha}$$
$$= \frac{1}{2}\alpha\frac{(1+\beta/\frac{1}{2}\alpha)}{u_{b0}\lambda_0 Z_0^{\beta/\frac{1}{2}\alpha}}\cdot(r/r_0)^{-(\beta/\alpha)(C_{\text{sh}}-M\sigma_{\text{D}}-1)-(M\sigma_{\text{D}}-\hat{C})+a_u}, \tag{B8}$$

leading to

$$\frac{d}{dr}\left[(r/r_0)^{(1+\beta/\frac{1}{2}\alpha)\frac{1}{2}(C_{\text{sh}}-M\sigma_{\text{D}}-1)}Z^{-1-\beta/\frac{1}{2}\alpha}\right] = \frac{1}{2}\alpha\frac{(1+\beta/\frac{1}{2}\alpha)}{u_{b0}\lambda_0 Z_0^{\beta/\frac{1}{2}\alpha}}\cdot(r/r_0)^{\frac{1}{2}(C_{\text{sh}}-3M\sigma_{\text{D}}-1)+\hat{C}+a_u}, \tag{B9}$$

Furthermore, we set

$$\varepsilon \equiv \frac{1}{2}(C_{\text{sh}} - 3M\sigma_{\text{D}} - 1) + \hat{C} + a_u + 1. \tag{B10}$$

Then, the radial dependence in right-hand side is written as $(r/r_0)^{-1+\varepsilon}$. Its integration is expanded as $\frac{1}{\varepsilon}[e^{\varepsilon\ln(r/r_0)}-1] \cong \ln(r/r_0) + \frac{1}{2}\varepsilon\ln^2(r/r_0)$, thus Eq.(B9) gives

$$Z = Z_0\cdot(r/r_0)^{\frac{1}{2}(C_{\text{sh}}-M\sigma_{\text{D}}-1)}\cdot\left\{1+(\frac{1}{2}\alpha+\beta)\frac{Z_0 r_0}{u_{b0}\lambda_0}\cdot\left[\ln(r/r_0)+\frac{1}{2}\varepsilon\ln^2(r/r_0)\right]\right\}^{-\frac{1}{1+\beta/\frac{1}{2}\alpha}}, \tag{B11a}$$

(because at $r=r_0$, $Z=Z_0$). Then, from Eq.(B6b) we solve for $\lambda$,

$$\lambda = \lambda_0\cdot(r/r_0)^{M\sigma_{\text{D}}-\hat{C}}\cdot\left\{1+(\frac{1}{2}\alpha+\beta)\frac{Z_0 r_0}{u_{b0}\lambda_0}\cdot\left[\ln(r/r_0)+\frac{1}{2}\varepsilon\ln^2(r/r_0)\right]\right\}^{\frac{\beta/\frac{1}{2}\alpha}{1+\beta/\frac{1}{2}\alpha}}. \tag{B11b}$$

The turbulent heating is given by

$$\frac{dE_t}{dr} = \frac{\frac{1}{2}\alpha Z^3}{u_b\lambda}, \tag{B12}$$

i.e.,

$$\frac{dE_t}{dr} = \frac{\frac{1}{2}\alpha Z_0^3}{u_{b0}\lambda_0}\cdot(r/r_0)^{\frac{3}{2}(C_{\text{sh}}-M\sigma_{\text{D}}-1)-M\sigma_{\text{D}}+\hat{C}+a_u}$$
$$\times\left\{1+(\frac{1}{2}\alpha+\beta)\frac{Z_0 r_0}{u_{b0}\lambda_0}\cdot\left[\ln(r/r_0)+\frac{1}{2}\varepsilon\ln^2(r/r_0)\right]\right\}^{-\frac{3+\beta/\frac{1}{2}\alpha}{1+\beta/\frac{1}{2}\alpha}}. \tag{B13}$$

Expanding in terms of $\ln(r/r_0)$, the heating is given by

$$\frac{dE_t}{dr} \cong \frac{\frac{1}{2}\alpha Z_0^3}{u_{b0}\lambda_0} \cdot (r/r_0)^{-(2+M\sigma_D-C_{sh}-\varepsilon)}$$

$$\times \begin{cases} 1-(\tfrac{3}{2}\alpha+\beta)\dfrac{Z_0 r_0}{u_{b0}\lambda_0}\cdot \ln(r/r_0) \\ +\tfrac{1}{2}(\tfrac{3}{2}\alpha+\beta)\left(\dfrac{Z_0 r_0}{u_{b0}\lambda_0}\right)\left[(\tfrac{3}{2}\alpha+\beta)\left(\dfrac{Z_0 r_0}{u_{b0}\lambda_0}\right)-\varepsilon\right]\cdot \ln^2(r/r_0) \\ +O(\varepsilon^2) \end{cases} \quad (B14)$$

where we used of $\varepsilon$ as defined in Eq.(B10). Then, ignoring terms $O[\ln^2(r/r_0)]$, we obtain

$$\frac{dE_t}{dr} \cong \frac{\frac{1}{2}\alpha Z_0^3}{u_{b0}\lambda_0}\cdot (r/r_0)^{-(2+M\sigma_D-C_{sh}-\varepsilon)}\cdot \left\{1-(\tfrac{3}{2}\alpha+\beta)\frac{Z_0 r_0}{u_{b0}\lambda_0}\cdot \ln(r/r_0)\right\}, \quad (B15)$$

which can be written as

$$\frac{dE_t}{dr} \cong A\cdot \left(\frac{r}{r_0}\right)^{-B}\cdot \left[1-C\cdot \ln\left(\frac{r}{r_0}\right)\right], \quad (B16a)$$

with

$$A = \frac{\frac{1}{2}\alpha Z_0^3}{\lambda_0 u_{b0}}, \quad B = 2+M\sigma_D-C_{sh}-\varepsilon, \quad C = \frac{(\tfrac{3}{2}\alpha+\beta)Z_0 r_0}{\lambda_0 u_{b0}} \quad (B16b)$$

Notes

(1) By comparing Eq.(B10) and Eq.(B15), we derive

$$B \cong 2+M\sigma_D-C_{sh}-\varepsilon \quad (B17)$$

(2) The arbitrary value of $\hat{C}$ is inherited to $\varepsilon$, and finally, to exponent $B$.